\documentclass[12pt,preprint]{aastex}

\usepackage{graphicx}
\usepackage{url}
\usepackage{amsmath}
\usepackage{epstopdf}
\usepackage{mathrsfs}
\usepackage{paralist}
\usepackage{IEEEtrantools}
\usepackage{subfigure}
\usepackage{color}

\shorttitle{The asteroseismic potential of {\it TESS}}
\shortauthors{Campante et al.}

\begin{document}

\title{The asteroseismic potential of {\it TESS}: exoplanet-host stars}

\author{T.~L.~Campante\altaffilmark{1,2}}
\email{campante@bison.ph.bham.ac.uk}
\and
\author{
M.~Schofield\altaffilmark{1,2},
J.~S.~Kuszlewicz\altaffilmark{1,2},
L.~Bouma\altaffilmark{3},
W.~J.~Chaplin\altaffilmark{1,2},
D.~Huber\altaffilmark{4,5,2},
J.~Christensen-Dalsgaard\altaffilmark{2},
H.~Kjeldsen\altaffilmark{2},
D.~Bossini\altaffilmark{1,2},
T.~S.~H.~North\altaffilmark{1,2},
T.~Appourchaux\altaffilmark{6},
D.~W.~Latham\altaffilmark{7},
J.~Pepper\altaffilmark{8},
G.~R.~Ricker\altaffilmark{3},
K.~G.~Stassun\altaffilmark{9,10},
R.~Vanderspek\altaffilmark{3},
J.~N.~Winn\altaffilmark{11,3}
}

\altaffiltext{1}{School of Physics and Astronomy, University of Birmingham, Edgbaston, Birmingham, B15 2TT, UK}
   
\altaffiltext{2}{Stellar Astrophysics Centre (SAC), Department of Physics and Astronomy, Aarhus University, Ny Munkegade 120, DK-8000 Aarhus C, Denmark}

\altaffiltext{3}{MIT Kavli Institute for Astrophysics and Space Research, 70 Vassar St., Cambridge, MA 02139, USA}

\altaffiltext{4}{Sydney Institute for Astronomy, School of Physics, University of Sydney, Sydney, Australia}

\altaffiltext{5}{SETI Institute, 189 Bernardo Avenue \#100, Mountain View, CA 94043, USA}

\altaffiltext{6}{Universit\'e Paris-Sud, Institut d'Astrophysique Spatiale, UMR 8617, CNRS, B\^atiment 121, 91405 Orsay Cedex, France}

\altaffiltext{7}{Harvard-Smithsonian Center for Astrophysics, 60 Garden Street, Cambridge, MA 02138, USA}

\altaffiltext{8}{Department of Physics, Lehigh University, Bethlehem, PA 18015, USA}

\altaffiltext{9}{Vanderbilt University, Physics \& Astronomy Department, 1807 Station B, Nashville, TN 37235, USA}

\altaffiltext{10}{Fisk University, Department of Physics, 1000 17th Ave.~N, Nashville, TN 37208, USA}

\altaffiltext{11}{Department of Physics, 77 Massachusetts Ave., Massachusetts Institute of Technology, Cambridge, MA 02139, USA}

\begin{abstract}
New insights on stellar evolution and stellar interiors physics are being made possible by asteroseismology. Throughout the course of the {\it Kepler} mission, asteroseismology has also played an important role in the characterization of exoplanet-host stars and their planetary systems. The upcoming NASA {\it Transiting Exoplanet Survey Satellite} ({\it TESS}) will be performing a near all-sky survey for planets that transit bright nearby stars. In addition, its excellent photometric precision, combined with its fine time sampling and long intervals of uninterrupted observations, will enable asteroseismology of solar-type and red-giant stars. Here we develop a simple test to estimate the detectability of solar-like oscillations in {\it TESS} photometry of any given star. Based on an all-sky stellar and planetary synthetic population, we go on to predict the asteroseismic yield of the {\it TESS} mission, placing emphasis on the yield of exoplanet-host stars for which we expect to detect solar-like oscillations. This is done for both the target stars (observed at a 2-min cadence) and the full-frame-image stars (observed at a 30-min cadence). A similar exercise is also conducted based on a compilation of known host stars. We predict that {\it TESS} will detect solar-like oscillations in a few dozen target hosts (mainly subgiant stars but also in a smaller number of F dwarfs), in up to 200 low-luminosity red-giant hosts, and in over 100 solar-type and red-giant known hosts, thereby leading to a threefold improvement in the asteroseismic yield of exoplanet-host stars when compared to {\it Kepler}'s.
\end{abstract}

\keywords{asteroseismology --- planets and satellites: detection --- space vehicles: instruments --- surveys --- techniques: photometric}

\section{Introduction}\label{sec:intro}
Asteroseismology is proving to be particularly relevant for the study of solar-type and red-giant stars \citep[for a review, see][and references therein]{ChaplinMiglioReview}, in great part due to the exquisite photometric data made available by the French-led {\it CoRoT} satellite \citep[COnvection ROtation and planetary Transits;][]{CoRoT}, NASA's {\it Kepler} space telescope \citep{Kepler} and, more recently, by the repurposed {\it K2} mission \citep{K2}. These stars exhibit solar-like oscillations, which are excited and intrinsically damped by turbulence in the outermost layers of a star's convective envelope. The information contained in solar-like oscillations allows fundamental stellar properties (e.g., mass, radius and age) to be precisely determined, while also allowing the internal stellar structure to be constrained to unprecedented levels, provided that individual oscillation mode parameters are measured. As a result, asteroseismology of solar-like oscillations is quickly maturing into a powerful tool whose impact is being felt more widely across different domains of astrophysics.

A noticeable example is the synergy between asteroseismology and exoplanetary science. Asteroseismology has been playing an important role in the characterization of exoplanet-host stars and their planetary systems, in particular over the course of the {\it Kepler} mission \citep[][]{HuberKOIs,Davies15,Kages}. Transit observations -- as carried out by {\it Kepler} -- are an indirect detection method, and are consequently only capable of providing planetary properties relative to the properties of the host star. The precise characterization of the host star through asteroseismology thus allows for inferences on the absolute properties of its planetary companions \citep[e.g.,][]{Kepler-36,Kepler-21,Kepler-37,Kepler-444,Gettel}. Moreover, information on the stellar inclination angle as provided by asteroseismology can lead to a better understanding of the planetary system dynamics and evolution \citep[e.g.,][]{ChaplinObliquities,Kepler-56,Campante16}. Another domain of application is that of orbital eccentricity determination based on the observed transit timescales \citep{SliskiKipping,VanEylenEccentricities}. Finally, the potential use of asteroseismology in measuring the levels of near-surface magnetic activity and in probing stellar activity cycles may help constrain the location of habitable zones around Sun-like stars.

The {\it Transiting Exoplanet Survey Satellite}\footnote{\url{http://tess.gsfc.nasa.gov/}} \citep[{\it TESS};][]{TESS1} is a NASA-sponsored Astrophysics Explorer mission that will perform a near all-sky survey for planets that transit bright nearby stars. Its launch is currently scheduled for December 2017. During the primary mission duration of two years, {\it TESS} will monitor the brightness of several hundred thousand main-sequence, low-mass stars over intervals ranging from one month to one year, depending mainly on a star's ecliptic latitude. Monitoring of these pre-selected target stars will be made at a cadence of {2\:{\rm min}}, while full-frame images will also be recorded every {30\:{\rm min}}. Being 10--100 times brighter than {\it Kepler} targets and distributed over a solid angle that is nearly 300 times larger, {\it TESS} host stars will be well suited for follow-up spectroscopy. \citet{TESS2} (hereafter S15) predicted the properties of the transiting planets detectable by {\it TESS} and of their host stars. {\it TESS} is expected to detect approximately 1700 transiting planets from $2\!\times\!10^5$ pre-selected target stars. The majority of the detected planets will have their radii in the sub-Neptune regime (i.e., 2--$4\,R_\earth$). Analysis of the full-frame images will lead to the additional detection of several thousand planets larger than $1.25\,R_\earth$ orbiting stars that are not among the pre-selected targets. 

Furthermore, {\it TESS}'s excellent photometric precision, combined with its fine time sampling and long intervals of uninterrupted observations, will enable asteroseismology of solar-type and red-giant stars, whose dominant oscillation periods range from several minutes to several hours. In this paper we aim at investigating the asteroseismic yield of the mission, placing emphasis on the yield of exoplanet-host stars for which we expect to detect solar-like oscillations. A broader study of the asteroseismic detections for stars that are not necessarily exoplanet hosts will be presented in a subsequent paper. The rest of the paper is organized as follows. A brief overview of {\it TESS} covering the mission design and survey operations is given in Sect.~\ref{sec:TESS}. \citet{ChaplinPredict} provide a simple recipe for estimating the detectability of solar-like oscillations in {\it Kepler} observations. In Sect.~\ref{sec:detect} we revisit that work and perform the necessary changes (plus a series of important updates) to make the recipe applicable to {\it TESS} photometry. Based on an existing all-sky stellar and planetary synthetic population, we then go on in Sect.~\ref{sec:simyield} to predict the yield of {\it TESS} exoplanet-host stars with detectable solar-like oscillations. A similar exercise is conducted in Sect.~\ref{sec:realyield}, although now based on a compilation of known (i.e., confirmed) host stars. We summarize and discuss our results in Sect.~\ref{sec:summary}.

\section{Overview of {\it TESS}}\label{sec:TESS}
Four identical cameras will be employed by {\it TESS}, each consisting of a lens assembly and a detector assembly with four $2048\!\times\!2048$ charge-coupled devices (CCDs). Each of the four lenses has an entrance pupil diameter of $10.5\:{\rm cm}$ and forms a $24\degr\!\times\!24\degr$ image on the four-CCD mosaic in its focal plane, hence leading to a pixel scale of $21\farcs1$. The effective collecting area of each camera is $69\:{\rm cm^2}$. The four camera fields are stacked vertically to create a combined field-of-view of $24\degr\!\times\!96\degr$ (or $2304\:{\rm sq.~deg.}$). 

\begin{figure}[!t]
\centering
\includegraphics[width=0.75\linewidth,trim={1cm 8cm 1cm 8cm},clip]{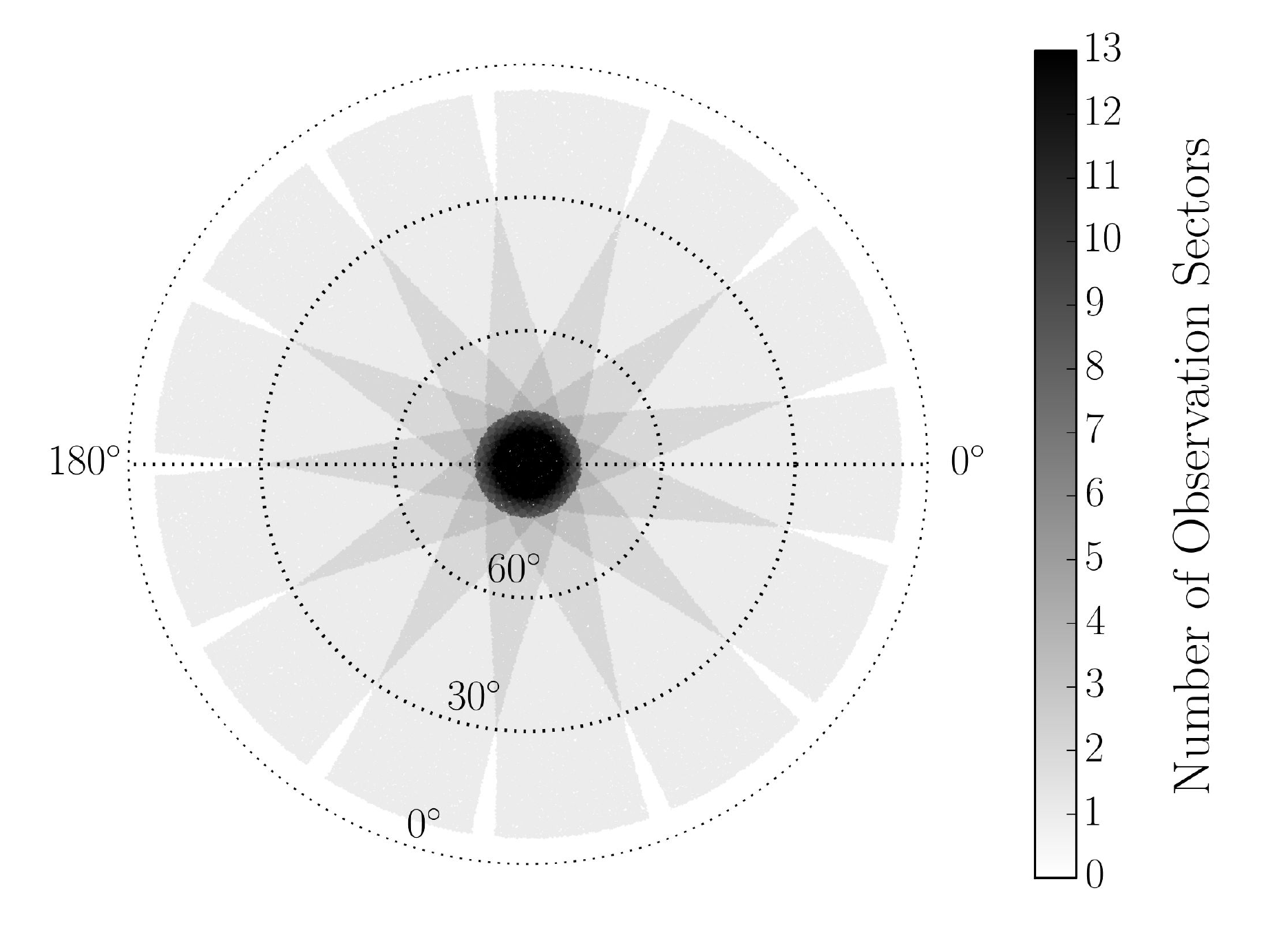}
\caption{\small Polar projection illustrating {\it TESS}'s coverage of a single ecliptic hemisphere.\label{fig:coverage}}
\end{figure}

{\it TESS} will observe from a thermally stable, low-radiation High Earth Orbit. {\it TESS}'s elliptical orbit will have a nominal perigee of $17\,R_\earth$ and a 13.7-day period in 2:1 resonance with the Moon's orbit. Over the course of the two-year duration of the primary mission, {\it TESS} will observe nearly the whole sky by dividing it into 26 observation sectors, 13 per ecliptic hemisphere. Each sector will be observed for 27.4 days (or two spacecraft orbits). Science operations will be interrupted at perigee for no more than 16 hours to allow for the downlink of the data, thus resulting in a high duty cycle of the observations. Figure \ref{fig:coverage} shows a polar projection illustrating the coverage of a single ecliptic hemisphere. The partially overlapping observation sectors are equally spaced in ecliptic longitude, extending from an ecliptic latitude of $6\degr$ to the ecliptic pole and beyond (the top camera is centered on the ecliptic pole). Successive sectors are positioned in order of increasing longitude (i.e., eastwardly), with the first pointing\footnote{This is the convention used in this work and in S15. The actual pointing coordinates will depend on the spacecraft's launch date.} centered at $0\degr$ of longitude. Approximately $30{,}000\:{\rm sq.~deg.}$ will be observed for at least 27.4 days. Moreover, observation sectors overlap near the ecliptic poles for increased sensitivity to smaller and longer-period planets in {\it James Webb Space Telescope}'s \citep[{\it JWST};][]{JWST} continuous viewing zone.

\begin{figure}[!t]
\centering
\includegraphics[width=0.95\linewidth]{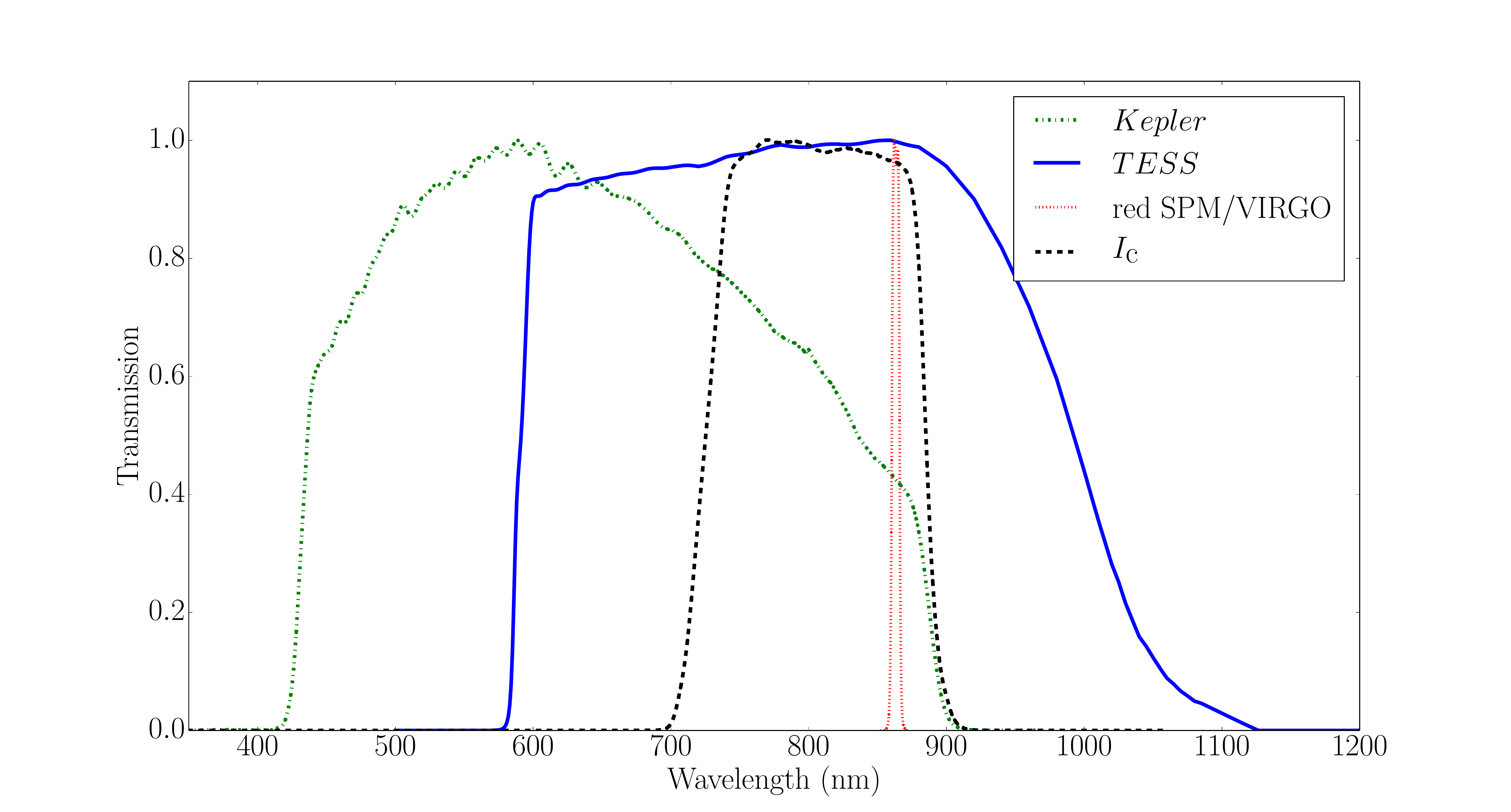}
\caption{\small {\it TESS} spectral response function. Also shown for comparison are the spectral response functions of {\it Kepler} and of the red channel of the SPM/VIRGO instrument on board {\it SoHO}, as well as the Johnson--Cousins $I_{\rm C}$ filter curve. Each curve has been normalized to have a maximum value of unity.\label{fig:bandpasses}}
\end{figure}

The {\it TESS} spectral response function is shown in Fig.~\ref{fig:bandpasses}. It is defined as the product of the long-pass filter transmission curve and the detector quantum efficiency curve. An enhanced sensitivity to red wavelengths is desirable, since cool red dwarfs will be preferentially targeted by {\it TESS} in the search for small transiting planets. The bandpass thus covers the range 600--$1000\:{\rm nm}$, being approximately centered on the Johnson--Cousins $I_{\rm C}$ band. The spectral response functions of {\it Kepler} and that of the red channel of the SPM/VIRGO instrument\footnote{The three-channel sun photometer (SPM) enables Sun-as-a-star helioseismology.} \citep{VIRGO} on board the {\it SoHO} spacecraft are also shown in Fig.~\ref{fig:bandpasses}.

New images will be acquired by each camera every 2 seconds. However, due to limitations in onboard data storage and telemetry, these 2-sec images will be stacked (before being downlinked to Earth) to produce two primary data products with longer effective exposure times: (i) subarrays of pixels centered on several hundred thousand pre-selected target stars will be stacked at a 2-min cadence, while (ii) full-frame images (FFIs) will be stacked every {30\:{\rm min}}. Up to $20{,}000$ 2-min-cadence slots (or the equivalent to $\sim\!10\,\%$ of the pre-selected target stars) will be allocated to the {\it TESS} Asteroseismic Science Consortium (TASC) over the course of the mission. In addition, a number of slots (notionally 1500) with faster-than-standard sampling, i.e., $20\:{\rm sec}$, will be reserved for the investigation of asteroseismic targets of special interest (mainly compact pulsators and main-sequence, low-mass stars).

A catalog of pre-selected target stars ($\ga2\!\times\!10^5$) will be monitored by {\it TESS} at a cadence of 2 min. This catalog will ideally include main-sequence stars that are sufficiently bright to maximize the prospects for detecting the transits of small planets (i.e., $R_{\rm p}<4\,R_\earth$). This leads to a limiting magnitude that will depend on spectral type, with $I_{\rm C}\la12$ for FGK dwarfs and $I_{\rm C}\la13$ for the smaller M dwarfs. In addition to the pre-selected targets, {\it TESS} will return FFIs with a cadence of 30 min, which will expand the search for transits to any sufficiently bright stars in the field of view that may have not been pre-selected. The longer integration time of the FFIs will, however, reduce the sensitivity to transits with a short duration. Over the course of the mission, the FFIs will be the source of precise photometry for approximately 20 million bright objects ($I_{\rm C}<14$--15).

\section{Predicting the detectability of solar-like oscillations}\label{sec:detect}
Solar-like oscillations are predominantly acoustic standing waves (or p modes). The oscillation modes are characterized by the radial order $n$ (related to the number of radial nodes), the spherical degree $l$ (specifying the number of nodal surface lines), and the azimuthal order $m$ (with $|m|$ specifying how many of the nodal surface lines cross the equator). Radial modes have $l\!=\!0$, whereas non-radial modes have $l\!>\!0$. Values of $m$ range from $-l$ to $l$, meaning that there are $2l+1$ azimuthal components for a given multiplet of degree $l$. Observed oscillation modes are typically high-order modes of low spherical degree, with the associated power spectrum showing a pattern of peaks with near-regular frequency separations. The most prominent separation is the large frequency separation, $\Delta \nu$, between neighboring overtones with the same spherical degree. The large frequency separation essentially scales as $\langle\rho\rangle^{1/2}$, where $\langle\rho\rangle\!\propto\!M/R^{3}$ is the mean density of a star with mass $M$ and radius $R$. Moreover, oscillation mode power is modulated by an envelope that generally assumes a bell-shaped appearance. The frequency at the peak of the power envelope is referred to as the frequency of maximum oscillation amplitude, $\nu_{\rm max}$. This frequency scales to very good approximation as $g\,T_{\rm eff}^{-1/2}$, where $g$ is the surface gravity and $T_{\rm eff}$ is the effective temperature. The fact that $\nu_{\rm max}$ mainly depends on $g$ makes it an indicator of the evolutionary state of a star.

\subsection{Detection test}\label{sec:detect_test}
In this work we adopt the test developed by \citet{ChaplinPredict} to estimate the detectability of solar-like oscillations in any given {\it Kepler} target, which looked for signatures of the bell-shaped power excess due to the oscillations \citep[see also][]{Campante14}. Below we revisit that work and detail the necessary changes (plus a series of important updates) to make the detection test applicable to {\it TESS} photometry.

{\bf Estimation of the detection probability, $p_{\rm detect}$.} The detection test is based upon the ratio of total mean mode power due to p-mode oscillations, $P_{\rm tot}$, to the total background power across the frequency range occupied by the oscillations, $B_{\rm tot}$. This quantity provides a global measure of the signal-to-noise ratio, S/N, in the oscillation spectrum, i.e.,
\begin{equation}
\label{eq:SNR_tot}
{\rm (S/N)_{tot}} = P_{\rm tot}/B_{\rm tot} \, .
\end{equation}
A total of $N$ independent frequency bins in the power spectrum enter the estimation of $P_{\rm tot}$ and $B_{\rm tot}$, and hence ${\rm (S/N)_{tot}}$:
\begin{equation}
N = W\,T \, ,               
\label{eq:bins}
\end{equation}
where
\begin{equation}
 W = \left\{ \,
    \begin{IEEEeqnarraybox}[][c]{l?s}
      \IEEEstrut
      1.32\,\nu_{\rm max}^{0.88} & if $\nu_{\rm max}\leq100\:{\rm \mu Hz}$ \, , \\
      \nu_{\rm max} & if $\nu_{\rm max}>100\:{\rm \mu Hz}$ \, .
      \IEEEstrut
    \end{IEEEeqnarraybox}
\right.
  \label{eq:envwidth}
\end{equation}
Here $T$ represents the length of the observations and is based on the maximum number of contiguous observation sectors for a given star. Moreover, we have assumed that the mode power is contained either within a range $\pm0.66\,\nu_{\rm max}^{0.88}$ \citep[][]{Mosser12} or $\pm\nu_{\rm max}/2$ \citep[][]{Stello07,Mosser10} around $\nu_{\rm max}$, with frequencies expressed in ${\rm \mu Hz}$. The width, $W$, of this range corresponds to twice the full width at half maximum of the power envelope (where a Gaussian-shaped envelope in frequency has been assumed). Note that any asymmetries of the power envelope have been disregarded.

When binning over $N$ bins, the statistics of the power spectrum of a pure noise signal is taken to be $\chi^2$ with $2N$ degrees of freedom \citep{Appourchaux04}. We begin by testing the null (or $H_0$) hypothesis that we observe pure noise. After specifying a false-alarm probability (or $p$-value) of $5\,\%$, we numerically compute the detection threshold ${\rm (S/N)_{\rm thresh}}$:
\begin{equation}
\label{eq:detect}
p=\int_x^\infty \frac{\exp(-x')}{\Gamma(N)}x'^{(N-1)}\,dx' \, ,
\end{equation}
where $x=1+{\rm (S/N)_{thresh}}$ and $\Gamma$ is the gamma function. Finally, the probability, $p_{\rm detect}$, that ${\rm (S/N)_{tot}}$ exceeds ${\rm (S/N)_{thresh}}$ is once more given by Eq.~(\ref{eq:detect}), but now setting $x=(1+{\rm (S/N)_{thresh}})/(1+{\rm (S/N)_{tot}})$. This last step can be thought of as testing the alternative (or $H_1$) hypothesis that we observe a signal embedded in noise. Throughout this work, we assume to be able to detect solar-like oscillations only in stars for which $p_{\rm detect}>0.5$. Next, we in turn detail how $P_{\rm tot}$ and $B_{\rm tot}$ are predicted.

{\bf Estimation of the total mean mode power, $P_{\rm tot}$.} The total mean mode power may be approximately predicted following:
\begin{equation}
P_{\rm tot} \approx 0.5\,c\,A_{\rm max}^2\,\eta^2(\nu_{\rm max})\,D^{-2}\,\frac{W}{\Delta\nu}\quad{\rm ppm^2} \, ,
\label{eq:Ptot}
\end{equation}
where $A_{\rm max}$ corresponds to the maximum oscillation amplitude of the radial ($l\!=\!0$) modes. The factor $c$ measures the effective number of p modes per order ($c\!=\!2.94$) and was computed following \citet{Bedding96} for a weighted wavelength of $797\:{\rm nm}$ representative of the {\it TESS} bandpass. We disregard the dependence of $c$ on $T_{\rm eff}$, $\log g$ and the metallicity, which could amount to relative variations of a few percent \citep{Ballot11}. The fraction in the above equation takes into account the contribution from all segments of width $\Delta\nu$ that fall in the range where mode power is present. On average, the power of the contributing segments will be $\sim\!0.5$ times that of the central segment, thus explaining the extra 0.5 factor in Eq.~(\ref{eq:Ptot}). The attenuation factor $\eta^2(\nu)$ takes into account the apodization of the oscillation signal due to the finite integration time. It is given by ${\rm sinc}^2\left[\pi/2\left(\frac{\nu}{\nu_{\rm Nyq}}\right)\right]$ for an integration duty cycle of $100\,\%$, where $\nu_{\rm Nyq}$ is the Nyquist frequency. Finally, a dilution (or wash-out) factor $D$ is introduced, which is defined as the ratio of the total flux in the photometric aperture from neighboring stars and the target star to the flux from the target star. This factor will be available for the simulated host stars introduced in Sect.~\ref{sec:simyield}, being otherwise set to $D\!=\!1$ (i.e., an isolated system).

The maximum oscillation amplitude, $A_{\rm max}$, is predicted based on:
\begin{equation}
\label{eq:Amax}
A_{\rm max}=(0.85)\,(2.5)\,\beta\,\left(\frac{R}{R_\sun}\right)^2\,\left(\frac{T_{\rm eff}}{T_{{\rm eff},\sun}}\right)^{0.5}\quad{\rm ppm} \, ,
\end{equation}
where
\begin{equation}
\label{eq:beta}
\beta = 1-\exp\left(-\frac{T_{\rm red}-T_{\rm eff}}{1550}\right)
\end{equation}
and
\begin{equation}
\label{eq:Tred}
T_{\rm red}\!=\!(8907)\,(L/{\rm L}_\sun)^{-0.093}\quad{\rm K} \, .
\end{equation}
Here and throughout we use $T_{{\rm eff},\sun}\!=\!5777\:{\rm K}$. Equation (\ref{eq:Amax}) is based on the prediction that the \textsc{rms} oscillation amplitude, $A_{\rm\textsc{rms}}$, observed in photometry at a wavelength $\lambda$, scales as $A_{\rm\textsc{rms}}\!\propto\!(L/M)^s/(\lambda T_{\rm eff}^r)$ \citep{KB95}, with $M$ subsequently eliminated using the scaling relation $M\!\propto\!T_{\rm eff}^{1.5}$ \citep[cf.][]{ChaplinPredict}. Accordingly, amplitudes are predicted to increase with increasing luminosity along the main sequence and relatively large amplitudes are expected for red giants. The exponent $s$ has been examined both theoretically and observationally, and found to lie in the range $0.7\!<\!s\!<\!1.5$ \cite[e.g.,][and references therein]{Corsaro13}. Here we adopt $s\!=\!1$ \citep[][]{ChaplinPredict}. The value of $r$ is chosen to be $r\!=\!2$ following a fit to observational data in \citet{KB95}. The factor $\beta$ is introduced to correct for the overestimation of oscillation amplitudes in the hottest solar-type stars, with the luminosity-dependent quantity $T_{\rm red}$ representing the temperature on the red edge of the radial-mode $\delta$ Scuti instability strip. The solar \textsc{rms} value $A_{{\rm max},\sun}$, as it would be measured by {\it Kepler}, is $A_{{\rm max},\sun}\!\sim\!2.5\:{\rm ppm}$. However, the absolute calibration of the predicted oscillation and granulation amplitudes depends on the spectral response of the instrument. {\it TESS} has a redder response than {\it Kepler} (cf.~Fig.~\ref{fig:bandpasses}), meaning observed amplitudes will be lower in the {\it TESS} data. Starting from the estimated {\it TESS} response, we followed the procedures outlined in \citet{Ballot11} to calculate a fractional multiplicative correction. We find that {\it TESS} oscillation (and granulation) amplitudes will be $\sim\!0.85$ times those observed with {\it Kepler}.

Even though Eq.~(\ref{eq:Amax}) has been calibrated based on solar-type stars alone \citep{ChaplinPredict}, it is also used here to predict the maximum oscillation amplitudes of red-giant stars. As a sanity check, we compared the red-giant oscillation amplitudes as predicted by Eq.~(\ref{eq:Amax}) with those obtained using the similar models $\mathcal{M}_1$ and $\mathcal{M}_{1,\beta}$ of \citet{Corsaro13}, whose calibration was based on over one thousand {\it Kepler} long-cadence targets. Having run such a test for a sequence of red-giant-branch (solar-calibrated) stellar models along a 1 ${\rm M}_\sun$ track, we obtained an ${\rm\textsc{rms}}$ relative difference of either $12\%$ (model $\mathcal{M}_1$) or $7\%$ (model $\mathcal{M}_{1,\beta}$). 

When predicting $A_{\rm max}$, the effect of stellar activity should be considered. Evidence has been found that high levels of stellar activity, tied to the magnetic field and rotation period of the star, tend to suppress the amplitudes of oscillation modes \citep{GarciaScience,ChaplinAct}. In order to incorporate an appropriate correction to the predicted mode amplitudes, the stellar activity levels must first be predicted from the fundamental stellar properties. This has, however, proven to be difficult, for a variety of reasons. The initial difficulty lies in describing how stellar activity can be measured from photometric time series. Throughout the {\it Kepler} mission, several activity proxies have been used \citep[e.g.,][]{Basri11,Campante14,Mathur14,Gilliland15} that show a high degree of correlation among them. However, predicting the absolute level of stellar activity remains a challenge. For instance, \citet{Gilliland11} attempted to predict stellar activity levels in {\it Kepler} stars by first predicting the chromospheric emission activity index $R'_{\rm HK}$, before converting this to a photometric measure. The prediction of $R'_{\rm HK}$ requires knowledge of the rotation period of the star, which can in principle be predicted from gyrochronology for low-mass stars ($M\!<\!1.3\,{\rm M}_\sun$) if the age of the star is also known \citep{Skumanich72,Aigrain04}. This is only applicable to main-sequence stars, since for more evolved stars the rotation period is no longer coupled to the stellar age in the same fashion. An additional problem with this procedure is that it in no way accounts for an activity cycle like the one observed in the Sun. Several challenges thus remain unsurmounted before stellar activity levels can be accounted for in the detection test and we ignore such a correction for the time being.

{\bf Estimation of the total background power, $B_{\rm tot}$.} The total background power is approximately given by
\begin{equation}
B_{\rm tot} \approx b_{\rm max}\,W \quad{\rm ppm^2} \, , 
\label{eq:Btot}
\end{equation}
where $b_{\rm max}$ is the background power spectral density from instrumental/shot noise and granulation at $\nu_{\rm max}$:
\begin{equation}
b_{\rm max} = b_{\rm instr} + P_{\rm gran}\quad{\rm ppm^2\,\mu Hz^{-1}} \, . 
\label{eq:bmax}
\end{equation}

\begin{figure}[!t]
\centering
\includegraphics[width=\linewidth]{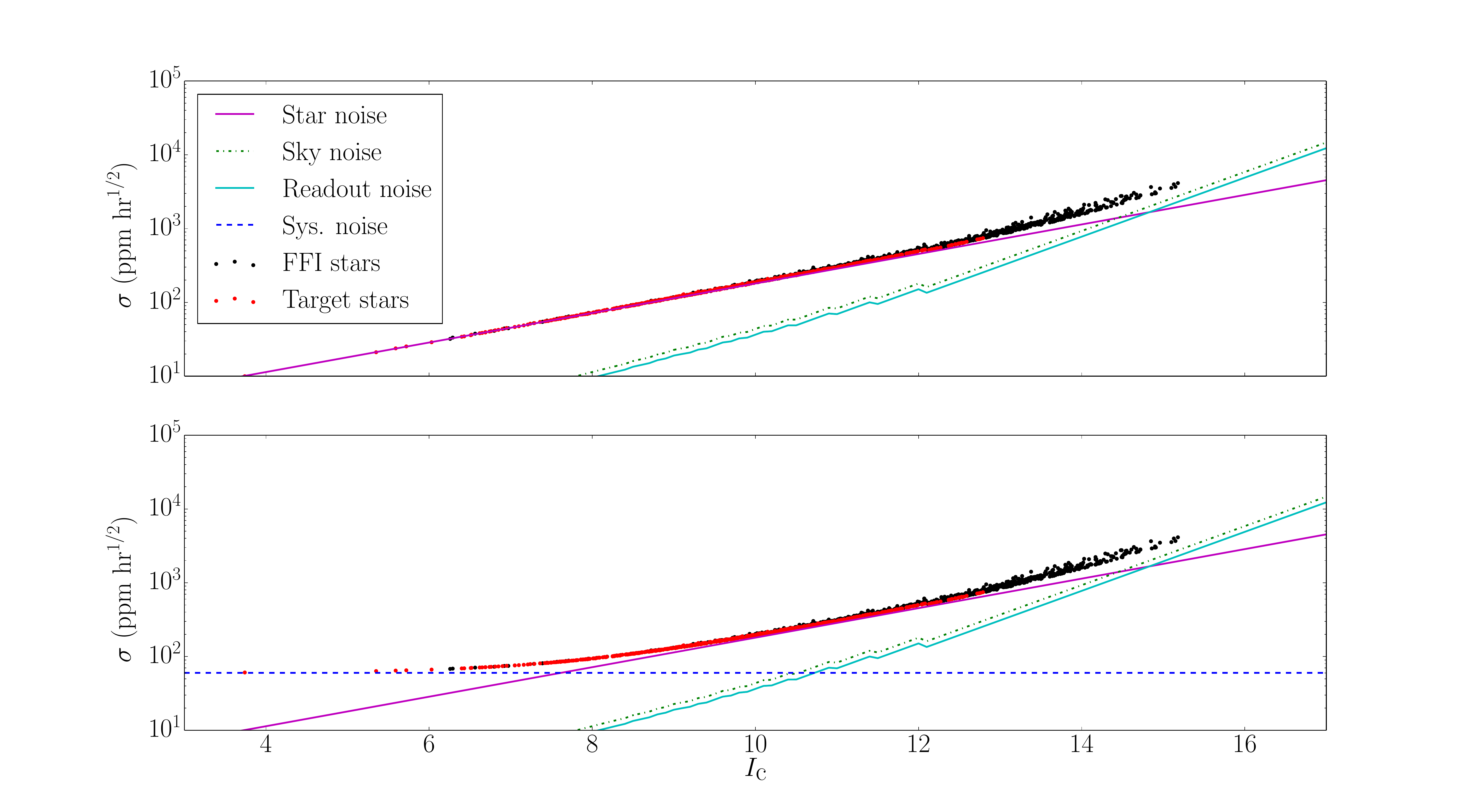}
\caption{\small Photometric noise model. Predicted \textsc{rms} noise, $\sigma$, per a 1-hour integration as a function of the apparent magnitude $I_{\rm C}$. The several noise components are represented by different line styles/colors. In the bottom panel a systematic noise level of $\sigma_{\rm sys}\!=\!60\:{\rm ppm\,hr^{1/2}}$ is assumed, while this systematic error term is absent from the top panel. The overall \textsc{rms} noise levels for a synthetic population (see Sect.~\ref{sec:simyield}) of host stars are also displayed (target stars in red and FFI stars in black).\label{fig:noisemodel}}
\end{figure}

The power spectral density due to instrumental/shot noise is given by \citep[e.g.,][]{Chaplin08}
\begin{equation}
\label{eq:binstr}
b_{\rm instr} = 2\times10^{-6}\,\sigma^2\,\Delta t\quad{\rm ppm^2\,\mu Hz^{-1}} \, ,
\end{equation}
where $\Delta t$ is the observational cadence. We use the photometric noise model for {\it TESS} presented in S15 to predict the \textsc{rms} noise, $\sigma$, per a given exposure time. This photometric noise model includes the photon-counting noise from the star (star noise), that from zodiacal light and background stars (sky noise), as well as the readout and systematic noise (instrumental noise). Figure \ref{fig:noisemodel} shows the contributions from the several noise components to the overall \textsc{rms} noise. The jagged appearance of the sky and readout noise components is due to the discretization of the number of pixels in the optimal photometric aperture. A systematic error term of $\sigma_{\rm sys}\!=\!60\:{\rm ppm\,hr^{1/2}}$ is included in the bottom panel of Fig.~\ref{fig:noisemodel}. This is an engineering requirement that is imposed on the design of the {\it TESS} photometer and not an estimate of the anticipated systematic noise level on 1-hour timescales. The systematic error term is assumed to scale with the total observing length as $T^{-1/2}$. It is perhaps unrealistic to assume that the systematic error will surpass $60\:{\rm ppm}$ for timescales shorter than one hour. Throughout this paper we will thus explore the implications of having $\sigma_{\rm sys}\!=\!0\:{\rm ppm\,hr^{1/2}}$ (ideal case) and $\sigma_{\rm sys}\!=\!60\:{\rm ppm\,hr^{1/2}}$ (regarded as a worst-case scenario).

Also shown in Fig.~\ref{fig:noisemodel} are the predicted \textsc{rms} noise levels for the simulated host stars of Sect.~\ref{sec:simyield} (target stars in red and FFI stars in black). The observed scatter is a result of the minute dependence of the overall noise on $T_{\rm eff}$ and a star's celestial coordinates. It can be seen that, for the brightest stars, the photometric precision is limited by the systematic noise floor (when present). We note that the central pixels of a stellar image will saturate for stars with $I_{\rm C}\la7.5$ during the 2-sec exposures, although high photometric precision is still expected down to $I_{\rm C}\!\approx\!4$ or brighter. For most of the stars in Fig.~\ref{fig:noisemodel}, whose magnitudes lie in the range $I_{\rm C}\!\approx\!7$--15, the photometric precision is instead dominated by stellar shot noise.

To model the granulation power spectral density, we adopt model F (with no mass dependence) of \citet{Kallinger14} and evaluate it at $\nu_{\rm max}$:
\begin{equation}
\label{eq:Pgran,real}
P_{\rm gran,real}(\nu_{\rm max}) = \eta^2(\nu_{\rm max})\,D^{-2}\,\sum_{i=1}^{2} \frac{(2\sqrt{2}/\pi)\,a_i^2/b_i}{1+(\nu_{\rm max}/b_i)^4}\quad{\rm ppm^2\,\mu Hz^{-1}} \, ,
\end{equation}
where the \textsc{rms} amplitude, $a_{1,2}$, and the characteristic frequencies, $b_1$ and $b_2$, are given by
\begin{subequations}
\begin{gather}
a_{1,2} = (0.85) (3382)\,\nu_{\rm max}^{-0.609}\quad{\rm ppm} \, , \label{eq:granpar_a}\\ 
b_1 = 0.317\,\nu_{\rm max}^{0.970}\quad{\rm \mu Hz} \, , \label{eq:granpar_b1}\\
b_2 = 0.948\,\nu_{\rm max}^{0.992}\quad{\rm \mu Hz} \, . \label{eq:granpar_b2}
\end{gather}
\end{subequations}
This model was found by \citet{Kallinger14} to be statistically preferred after a Bayesian model comparison that considered different approaches to quantifying the signature of stellar granulation. The model consists of two super-Lorentzian functions representing separate classes of physical processes such as stellar activity and/or different granulation scales. Model parameters have been calibrated via fits to the power spectra of a large set of {\it Kepler} targets, hence explaining the 0.85 multiplicative correction in Eq.~(\ref{eq:granpar_a}) to convert to {\it TESS} granulation amplitudes.

When a continuous signal is being sampled that contains frequency components above the Nyquist frequency, $\nu_{\rm Nyq}\!\equiv\!1/(2\Delta t)$, these will give rise to an effect known as aliasing and the signal is then said to be undersampled. The aliased granulation power at $\nu_{\rm max}$, $P_{\rm gran,aliased}(\nu_{\rm max})$, is given by\footnote{Note that although $P_{\rm gran,real}(\nu'_{\rm max})$ is computed at $\nu'_{\rm max}$, the coefficients $a_{1,2}$ and $b_{1,2}$ are evaluated at $\nu_{\rm max}$.}
\begin{equation}
\label{eq:Pgran,aliased}
P_{\rm gran,aliased}(\nu_{\rm max}) \equiv P_{\rm gran,real}(\nu'_{\rm max}) \, ,
\end{equation}
with the folded frequency $\nu'_{\rm max}$ defined as
\begin{equation}
 \nu'_{\rm max} = \left\{ \,
    \begin{IEEEeqnarraybox}[][c]{l?s}
      \IEEEstrut
      \nu_{\rm Nyq} + (\nu_{\rm Nyq} - \nu_{\rm max}) & if $\nu_{\rm max}\leq\nu_{\rm Nyq}$\, , \\
      \nu_{\rm Nyq} - (\nu_{\rm max} - \nu_{\rm Nyq}) & if $\nu_{\rm Nyq}<\nu_{\rm max}\leq2\nu_{\rm Nyq}$\, ,
      \IEEEstrut
    \end{IEEEeqnarraybox}
\right.
  \label{eq:aliasfreq}
\end{equation}
where we restrict ourselves to the range $[0,2\,\nu_{\rm Nyq}]$. The total granulation power spectral density (at $\nu_{\rm max}$) is then given by
\begin{equation}
\label{eq:Pgran}
P_{\rm gran} = P_{\rm gran,real}(\nu_{\rm max}) + P_{\rm gran,aliased}(\nu_{\rm max}) \, .
\end{equation}
The formalism above allows us to correctly predict the detectability of solar-like oscillations both in stars with $\nu_{\rm max}$ in the sub- ($\nu_{\rm max}\leq\nu_{\rm Nyq}$) and super-Nyquist ($\nu_{\rm Nyq}<\nu_{\rm max}\leq2\nu_{\rm Nyq}$) regimes. The latter regime is particularly relevant for stars in FFIs \citep[cf.][]{superNyq}, for which $\nu_{\rm Nyq,FFI}\!\sim\!278\:{\rm \mu Hz}$, although not as much for target stars, since we do not expect to detect solar-like oscillations with $\nu_{\rm max}$ above $\nu_{\rm Nyq,target}\!\sim\!4167\:{\rm \mu Hz}$. 

\begin{figure}[!t]
\centering
\includegraphics[width=\linewidth]{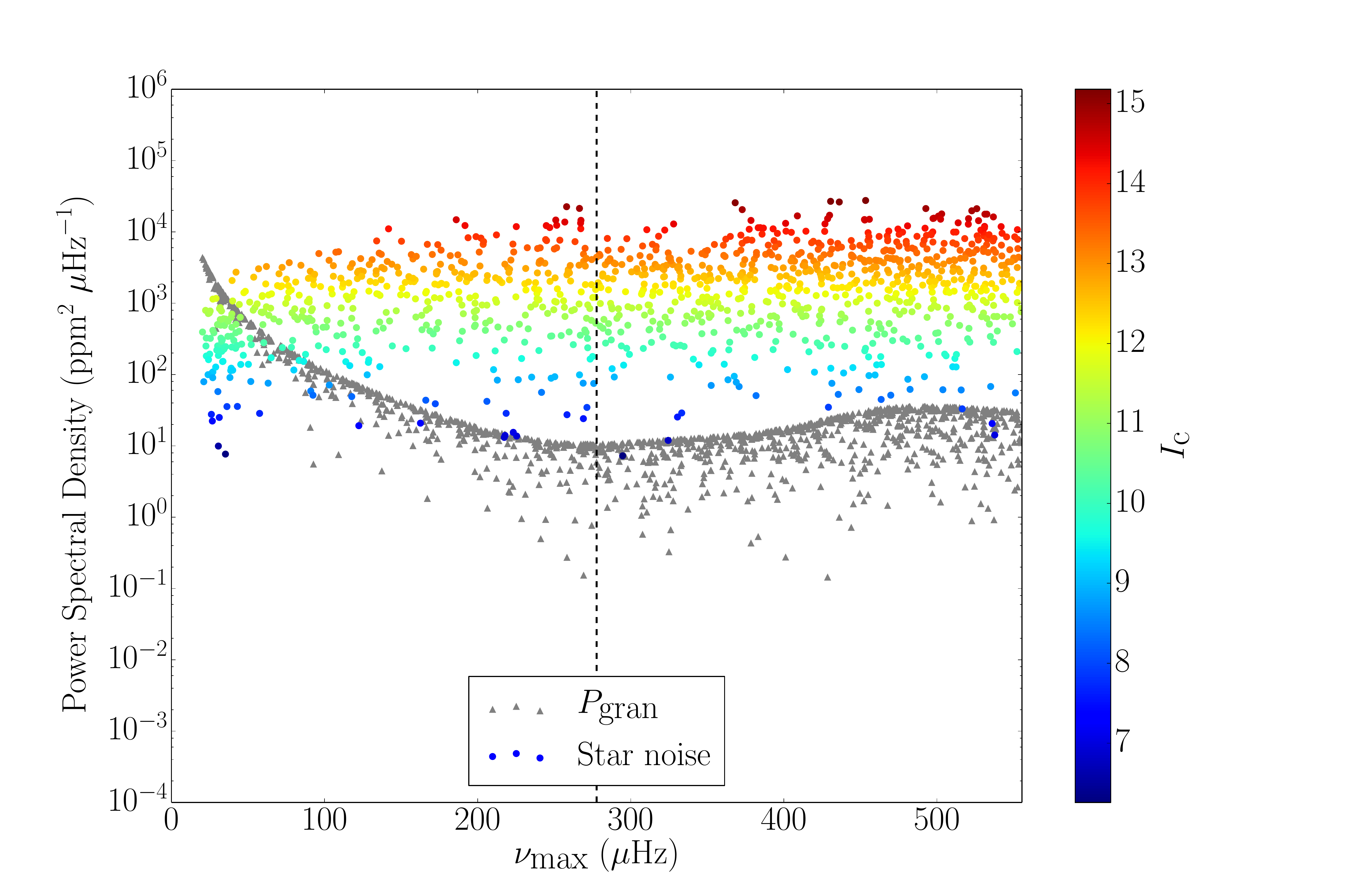}
\caption{\small Contributions from granulation ($P_{\rm gran}$) and stellar shot noise to the background power spectral density of a synthetic population (see Sect.~\ref{sec:simyield30min}) of FFI host stars. The contribution from stellar shot noise is color-coded according to $I_{\rm C}$. The vertical dashed line represents $\nu_{\rm Nyq,FFI}\!\sim\!278\:{\rm \mu Hz}$.\label{fig:Btot}}
\end{figure}

Figure \ref{fig:Btot} shows the contributions from granulation ($P_{\rm gran}$) and stellar shot noise to the background power spectral density (Eq.~\ref{eq:bmax}) of the simulated FFI host stars in Sect.~\ref{sec:simyield30min}. The observed scatter for $P_{\rm gran}$ is entirely due to the varying dilution factor, $D$. Stellar shot noise is seen to dominate over granulation across most of the plotted frequency range. This is in stark contrast to what was observed with {\it Kepler} photometry \citep[e.g.,][]{Mathur11,Karoff13,Kallinger14} and is mostly due to the smaller (by a factor of $\sim\!10^2$) effective collecting area of the individual {\it TESS} cameras. While this will likely make robust modeling of the granulation profile a challenge, it does not necessarily mean that oscillations cannot be detected, as shown below. 

{\bf Estimation of $\nu_{\rm max}$ and $\Delta\nu$.} The values of $\nu_{\rm max}$ and $\Delta\nu$ used as input in the detection test are predicted from the stellar mass (when available; cf.~Sect.~\ref{sec:detect_HR}), stellar radius and effective temperature according to the scaling relations \cite[e.g.,][and references therein]{Kallinger10}:
\begin{equation}
\nu_{\rm max} = \nu_{{\rm max},\sun} \left(\frac{M}{{\rm M}_\sun}\right) \left(\frac{R}{R_\sun}\right)^{-2} \left(\frac{T_{\rm eff}}{T_{{\rm eff},\sun}}\right)^{-0.5}
\label{eq:scaling_numax}
\end{equation}
and
\begin{equation}
\Delta\nu = \Delta\nu_{\sun} \left(\frac{M}{{\rm M}_\sun}\right)^{0.5} \left(\frac{R}{R_\sun}\right)^{-1.5} \, ,
\label{eq:scaling_Deltanu}
\end{equation}
with $\nu_{{\rm max},\sun}\!=\!3090\:{\rm \mu Hz}$ and $\Delta\nu_\sun\!=\!135.1\:{\rm \mu Hz}$. If no stellar mass is available (cf.~Sects.~\ref{sec:simyield} and \ref{sec:realyield}), we then eliminate $M$ from Eqs.~(\ref{eq:scaling_numax}) and (\ref{eq:scaling_Deltanu}) using the relation \citep{Stello09}
\begin{equation}
\label{eq:Deltanu_numax}
\Delta\nu \propto \nu_{\rm max}^{0.77} \, ,
\end{equation}
whose calibration was based on a cohort of stars with $\nu_{\rm max}$ in the range $15\!\la\!\nu_{\rm max}\!\la\!4500\:{\rm \mu Hz}$. We note that the exponent in the previous equation varies slightly depending on the range in $\nu_{\rm max}$ being considered \citep[][]{Huber11}. However, for the purpose of this work, the use of a `unified' relation such as Eq.~(\ref{eq:Deltanu_numax}) seems justified. The resulting scaling relations for $\nu_{\rm max}$ and $\Delta\nu$ in terms of the stellar radius and effective temperature are:
\begin{equation}
\nu_{\rm max} = \nu_{{\rm max},\sun} \left(\frac{R}{R_\sun}\right)^{-1.85} \left(\frac{T_{\rm eff}}{T_{{\rm eff},\sun}}\right)^{0.92}
\label{eq:scaling_numax_ii}
\end{equation}
and
\begin{equation}
\Delta\nu = \Delta\nu_{\sun} \left(\frac{R}{R_\sun}\right)^{-1.42} \left(\frac{T_{\rm eff}}{T_{{\rm eff},\sun}}\right)^{0.71} \, .
\label{eq:scaling_Deltanu_ii}
\end{equation}
As a sanity check, we compared the output values from Eqs.~(\ref{eq:scaling_numax_ii}) and (\ref{eq:scaling_Deltanu_ii}) with those from Eqs.~(\ref{eq:scaling_numax}) and (\ref{eq:scaling_Deltanu}) across the full $\nu_{\rm max}$ and $\Delta\nu$ ranges. Based on a sequence of (solar-calibrated) stellar models along a 1 ${\rm M}_\sun$ track, we obtained an ${\rm\textsc{rms}}$ relative difference of $3.9\%$ for $\nu_{\rm max}$ and $1.8\%$ for $\Delta\nu$, commensurate with typical fractional uncertainties measured by {\it Kepler} for these global parameters \citep[e.g.,][]{Kallinger10,Chaplin14}.

\subsection{Detectability of solar-like oscillations across the H--R diagram}\label{sec:detect_HR}
\begin{figure}[!t]
\centering
  \subfigure[$T=27\:{\rm d}$, $\sigma_{\rm sys}\!=\!0\:{\rm ppm\,hr^{1/2}}$.]{%
  \includegraphics[width=.5\textwidth]{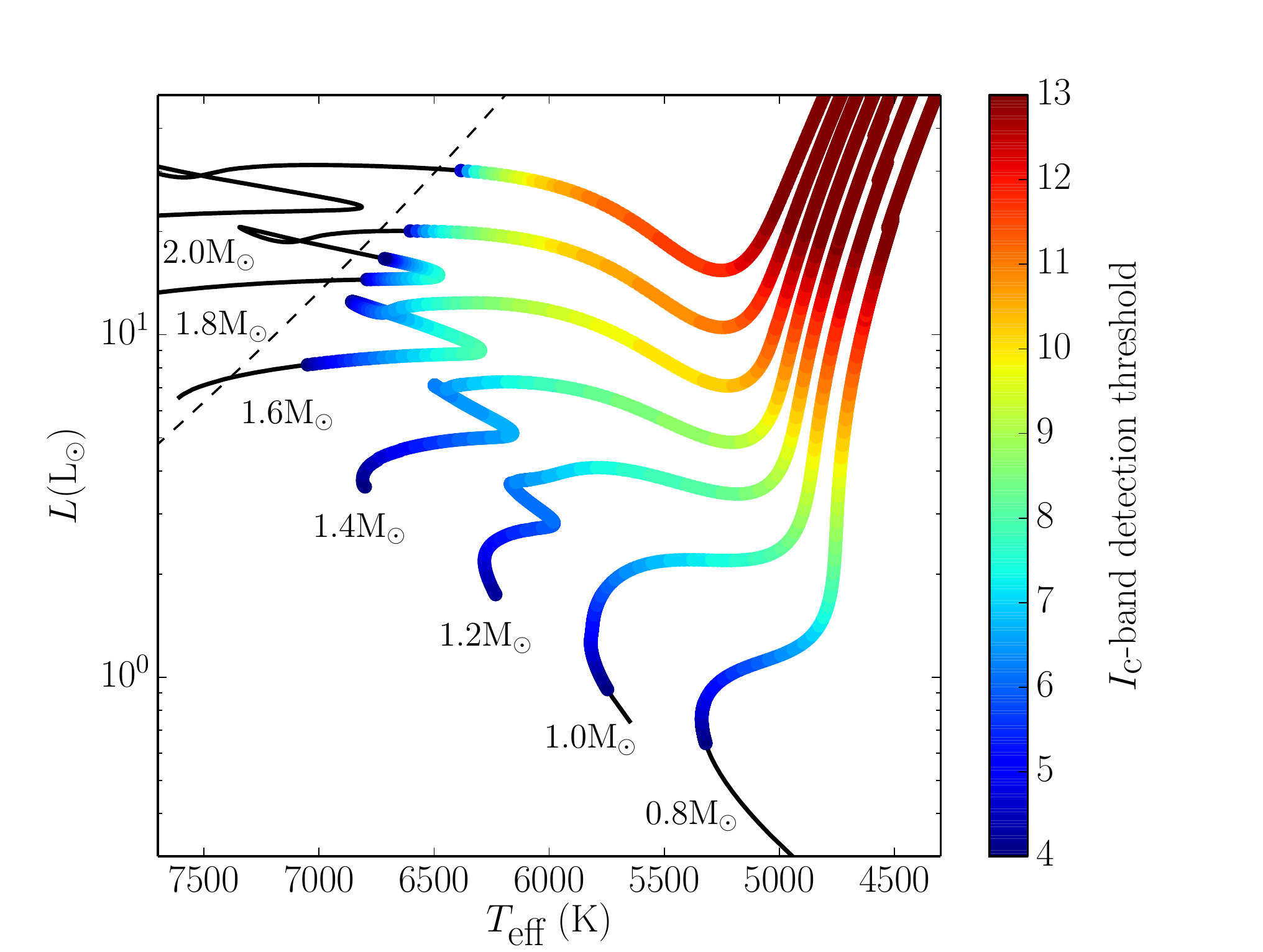}}\hfill
  \subfigure[$T=27\:{\rm d}$, $\sigma_{\rm sys}\!=\!60\:{\rm ppm\,hr^{1/2}}$.]{%
  \includegraphics[width=.5\textwidth]{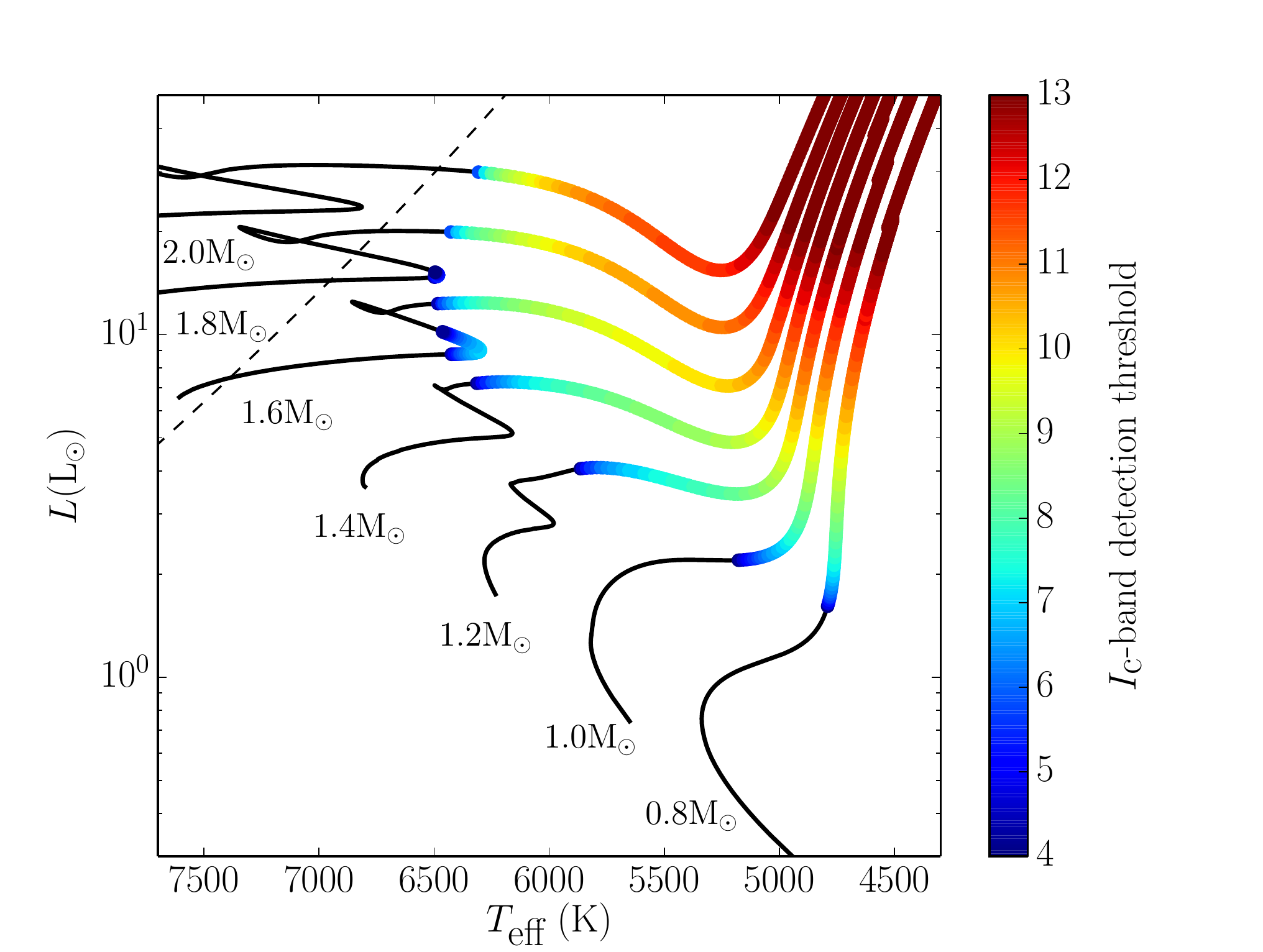}}\\
  \subfigure[$T=351\:{\rm d}$, $\sigma_{\rm sys}\!=\!0\:{\rm ppm\,hr^{1/2}}$.]{%
  \includegraphics[width=.5\textwidth]{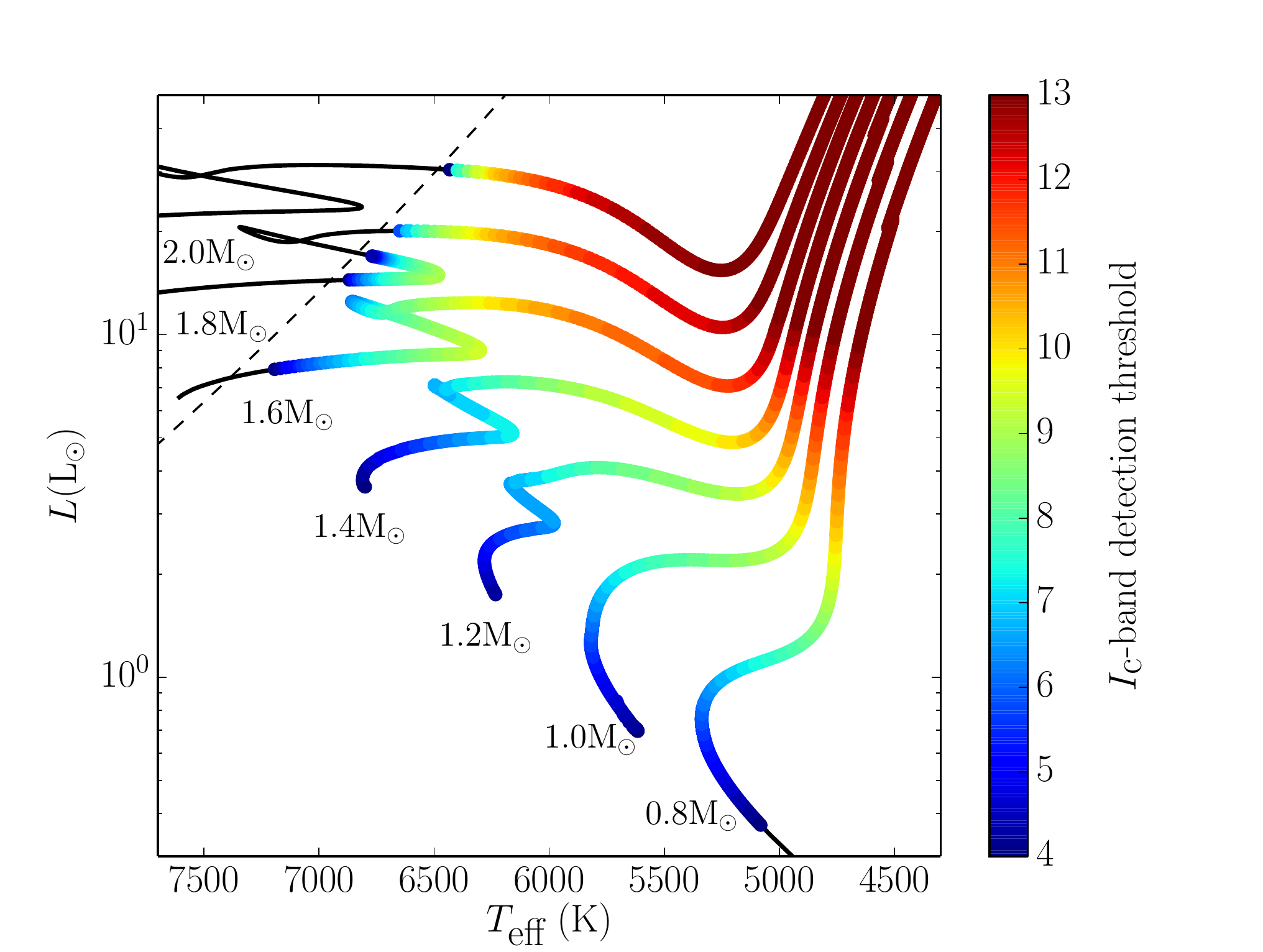}}\hfill
  \subfigure[$T=351\:{\rm d}$, $\sigma_{\rm sys}\!=\!60\:{\rm ppm\,hr^{1/2}}$.]{%
  \includegraphics[width=.5\textwidth]{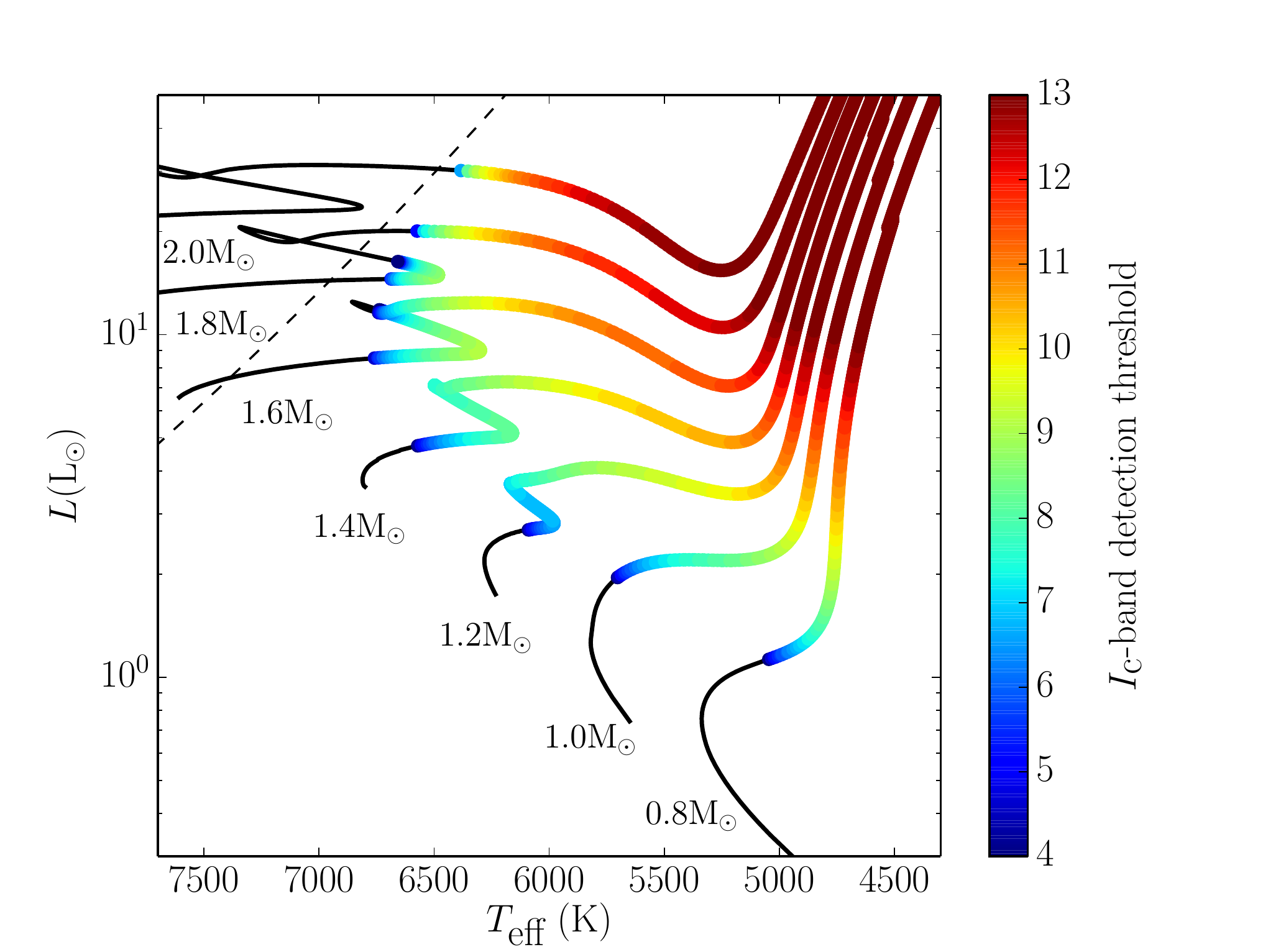}}
  \caption{\small Detectability of solar-like oscillations with {\it TESS} across the H--R diagram for a cadence of $\Delta t\!=\!2\:{\rm min}$. Solar-calibrated evolutionary tracks spanning the mass range 0.8--$2.0\,{\rm M}_\sun$ (in steps of $0.2\,{\rm M}_\sun$) are displayed. $I_{\rm C}$-band detection thresholds are color-coded (no detection is possible along those portions of the tracks shown as a thin black line). Modeled stars were assumed to be isolated (i.e., $D\!=\!1$). The slanted dashed line represents the red edge of the $\delta$ Scuti instability strip. The several panels consider different combinations of the length of the observations ($T$) and systematic noise level ($\sigma_{\rm sys}$), as indicated.}\label{fig:stellar_tracks_120s_lum}
\end{figure}

\begin{figure}[!t]
\centering
  \subfigure[$T=27\:{\rm d}$, $\sigma_{\rm sys}\!=\!60\:{\rm ppm\,hr^{1/2}}$.]{%
  \includegraphics[width=.5\textwidth]{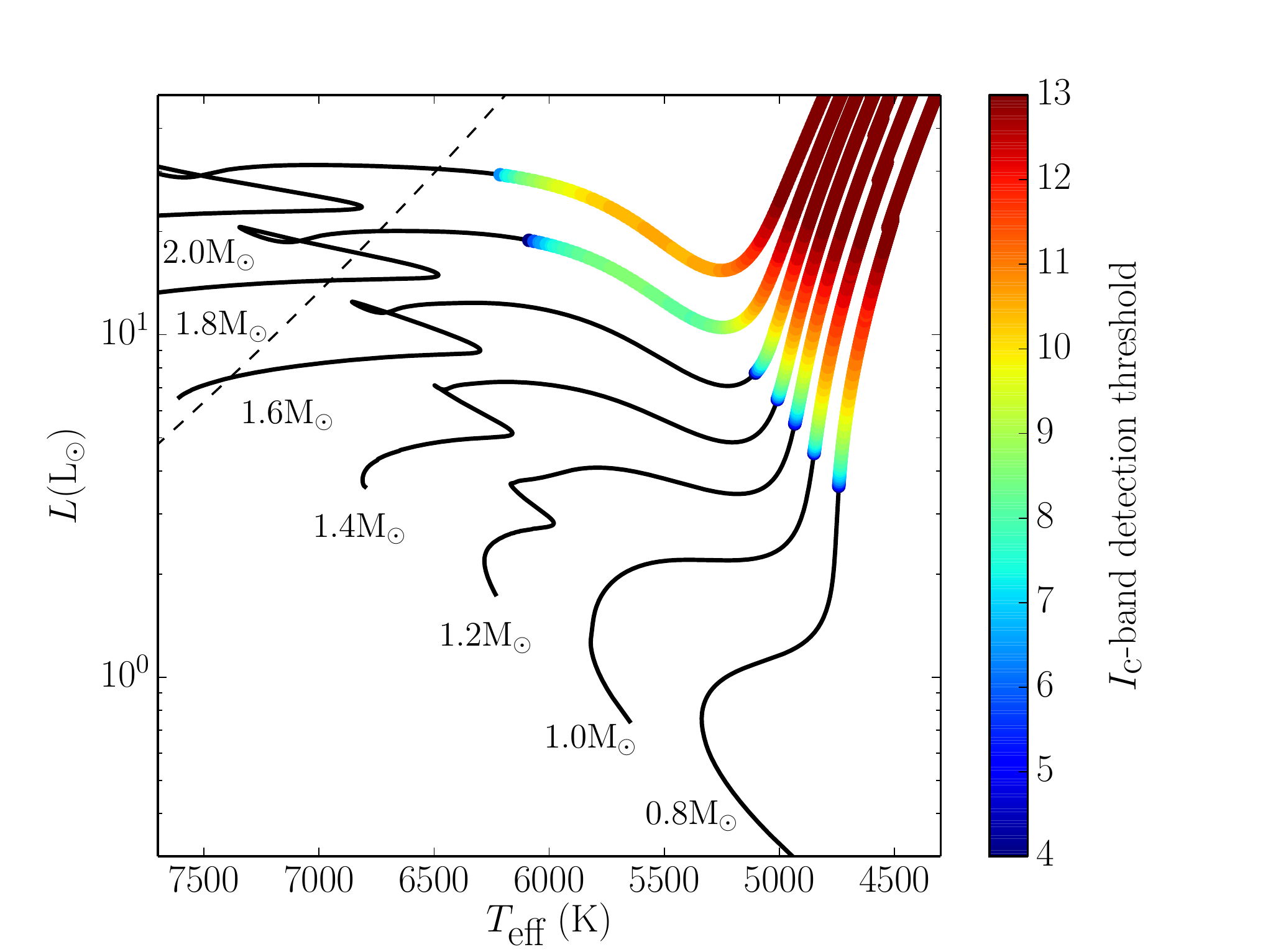}}\hfill
  \subfigure[$T=351\:{\rm d}$, $\sigma_{\rm sys}\!=\!60\:{\rm ppm\,hr^{1/2}}$.]{%
  \includegraphics[width=.5\textwidth]{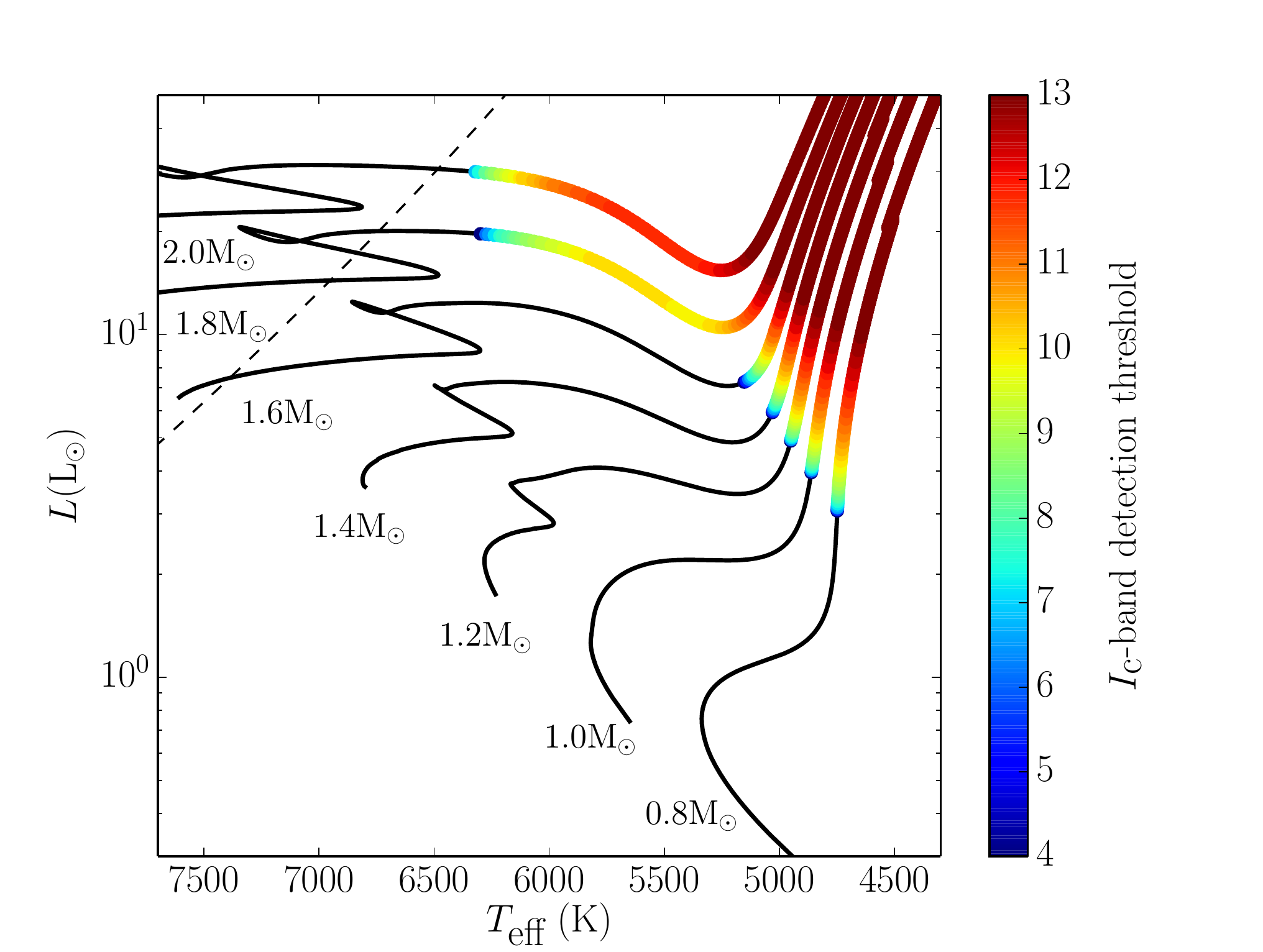}}
  \caption{\small Detectability of solar-like oscillations with {\it TESS} across the H--R diagram for a cadence of $\Delta t\!=\!30\:{\rm min}$. Solar-calibrated evolutionary tracks spanning the mass range 0.8--$2.0\,{\rm M}_\sun$ (in steps of $0.2\,{\rm M}_\sun$) are displayed. $I_{\rm C}$-band detection thresholds are color-coded (no detection is possible along those portions of the tracks shown as a thin black line). Modeled stars were assumed to be isolated (i.e., $D\!=\!1$). The slanted dashed line represents the red edge of the $\delta$ Scuti instability strip. The two panels consider different lengths of the observations ($T$) and a systematic noise level of $\sigma_{\rm sys}\!=\!60\:{\rm ppm\,hr^{1/2}}$, as indicated.}\label{fig:stellar_tracks_1800s_lum}
\end{figure}

\begin{figure}[!t]
\centering
  \subfigure[$T=27\:{\rm d}$, $\sigma_{\rm sys}\!=\!60\:{\rm ppm\,hr^{1/2}}$.]{%
  \includegraphics[width=.5\textwidth]{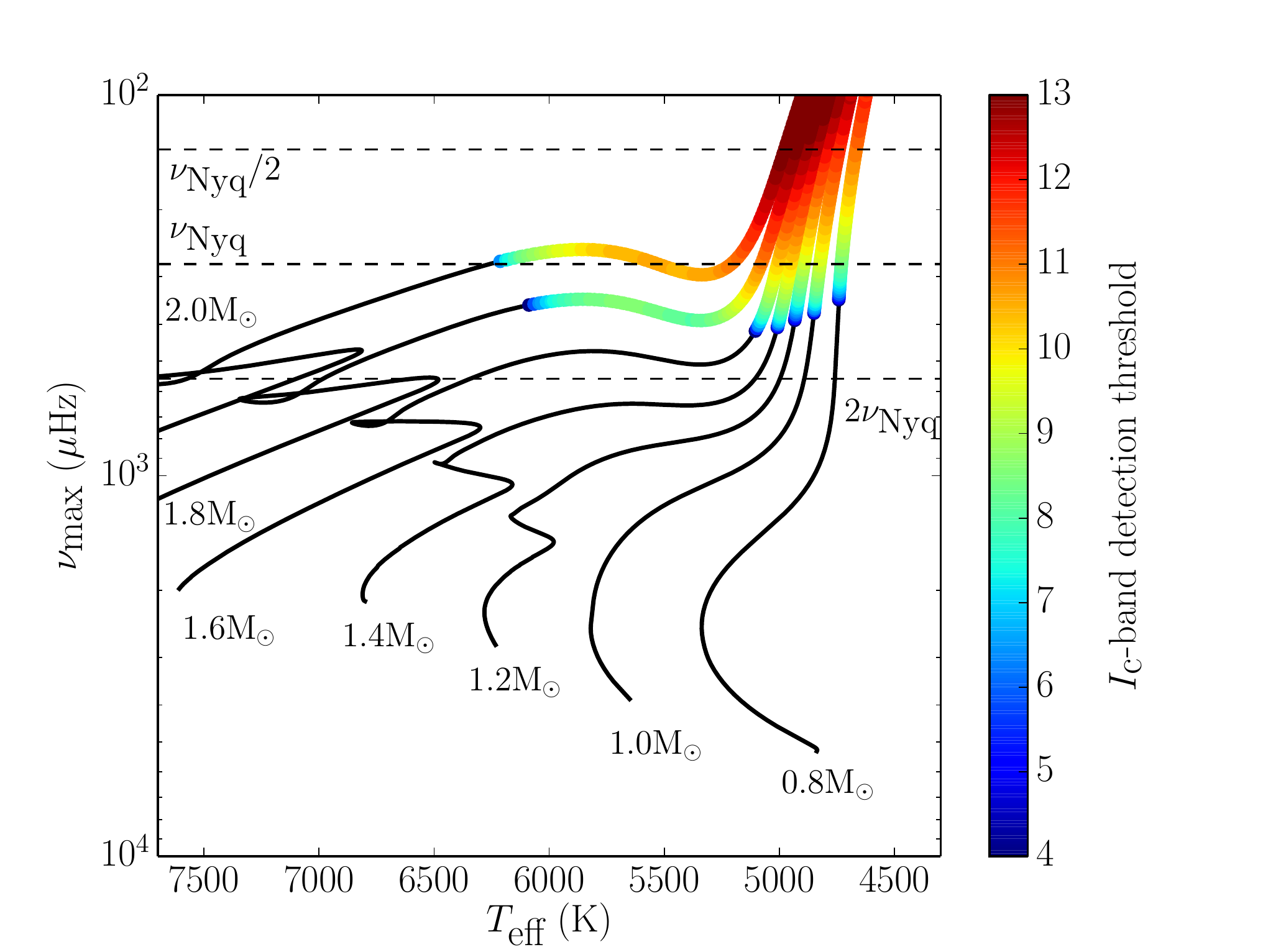}}\hfill
  \subfigure[$T=351\:{\rm d}$, $\sigma_{\rm sys}\!=\!60\:{\rm ppm\,hr^{1/2}}$.]{%
  \includegraphics[width=.5\textwidth]{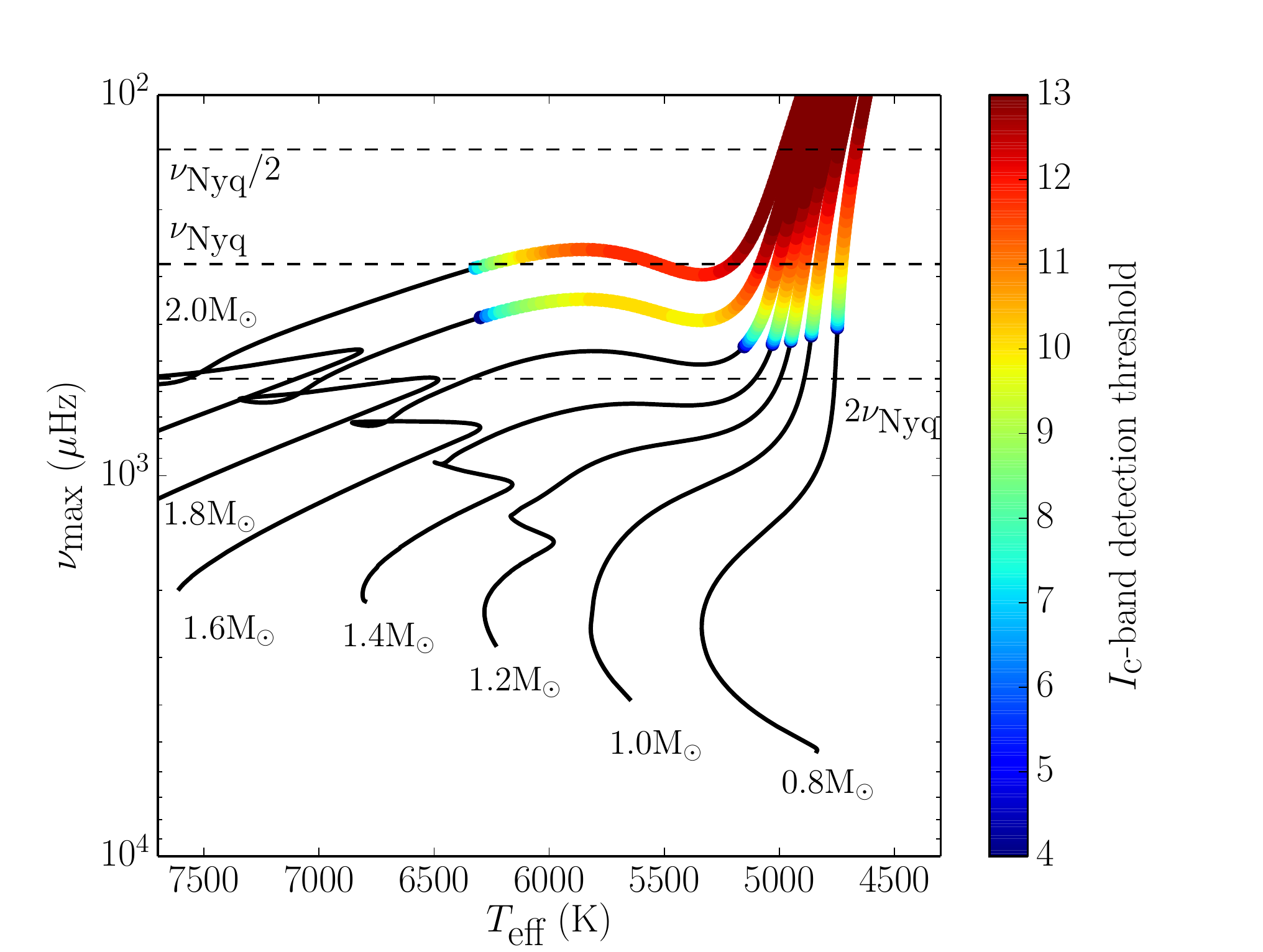}}
  \caption{\small Detectability of solar-like oscillations with {\it TESS} across an asteroseismic H--R diagram for a cadence of $\Delta t\!=\!30\:{\rm min}$. Note that $\nu_{\rm max}$ is now plotted along the vertical axis and not luminosity. Horizontal dashed lines indicate $\nu_{\rm Nyq}/2$, $\nu_{\rm Nyq}$ and $2\nu_{\rm Nyq}$. Solar-calibrated evolutionary tracks spanning the mass range 0.8--$2.0\,{\rm M}_\sun$ (in steps of $0.2\,{\rm M}_\sun$) are displayed. $I_{\rm C}$-band detection thresholds are color-coded (no detection is possible along those portions of the tracks shown as a thin black line). Modeled stars were assumed to be isolated (i.e., $D\!=\!1$). The two panels consider different lengths of the observations ($T$) and a systematic noise level of $\sigma_{\rm sys}\!=\!60\:{\rm ppm\,hr^{1/2}}$, as indicated.}\label{fig:stellar_tracks_1800s_numax}
\end{figure}

Figures \ref{fig:stellar_tracks_120s_lum}--\ref{fig:stellar_tracks_1800s_numax} depict the detectability of solar-like oscillations with {\it TESS} across the Hertzsprung--Russell (H--R) diagram. We focus on that portion of the H--R diagram populated by solar-type and low-luminosity red-giant stars (i.e., up to the red-giant branch bump), bound at high effective temperatures by the red edge of the $\delta$ Scuti instability strip. The detection code was applied along several solar-calibrated stellar-model tracks spanning the mass range 0.8--$2.0\,{\rm M}_\sun$ (in steps of $0.2\,{\rm M}_\sun$). These stellar models were computed using the Modules for Experiments in Stellar Astrophysics \citep[MESA;][]{Paxton11,Paxton13} evolution code.

In Figure \ref{fig:stellar_tracks_120s_lum} we consider two different observing lengths (corresponding to 1 and 13 observation sectors) and a cadence of $\Delta t\!=\!2\:{\rm min}$. Further assuming a systematic noise level of $\sigma_{\rm sys}\!=\!60\:{\rm ppm\,hr^{1/2}}$, detection of solar-like oscillations in main-sequence stars will not be possible for $T\!=\!27\:{\rm d}$. Increasing the observing length to $T\!=\!351\:{\rm d}$ (relevant for stars near the ecliptic poles) may lead to the marginal detection of oscillations in (very bright) main-sequence stars more massive than the Sun. In both cases, detection of oscillations in subgiant and red-giant stars is nonetheless made possible, owing to their higher intrinsic amplitudes. As one would expect, this situation is significantly improved as the systematic noise level is brought down to $\sigma_{\rm sys}\!=\!0\:{\rm ppm\,hr^{1/2}}$, with detections now being made possible for the brightest main-sequence stars over a range of masses. The longer 30-min cadence is considered in Figs.~\ref{fig:stellar_tracks_1800s_lum} and \ref{fig:stellar_tracks_1800s_numax}, where we have assumed a systematic noise level of $\sigma_{\rm sys}\!=\!60\:{\rm ppm\,hr^{1/2}}$ only. FFIs will allow detecting oscillations in red-giant stars down to relatively faint magnitudes. Furthermore, it becomes apparent from Fig.~\ref{fig:stellar_tracks_1800s_numax} that it should be possible to detect oscillations in the super-Nyquist regime for the brightest red giants.

\section{Asteroseismic yield based on simulated data}\label{sec:simyield}
In S15 the authors predicted the properties of the transiting planets detectable by {\it TESS} and of their host stars, having done so for both the cohorts of target and FFI systems. Predictions were also made of the population of eclipsing binary stars that produce false-positive photometric signals. These predictions are based on a Monte Carlo simulation of a population of nearby stars generated using the TRIdimensional modeL of thE GALaxy \citep[TRILEGAL;][]{TRILEGAL} population synthesis code. Any simulated star that could be searched for transiting planets is included in a so-called `bright catalog' (with 2MASS $K_S$ magnitude $K_S\!<\!15$) containing $1.58\!\times\!10^8$ stars. The $2\!\times\!10^5$ target stars are then selected from this catalog. The simulation employs planet occurrence rates derived from {\it Kepler} \citep{Fressin13,DressChar} whose completeness is high for the planetary periods and radii relevant to {\it TESS}, and a model for the photometric performance of the {\it TESS} cameras. In the present section, we apply the detection test to the synthetic population of host stars obtained in this way in order to predict the yield of {\it TESS} hosts with detectable solar-like oscillations.

\subsection{{\it TESS} target hosts}\label{sec:simyield2min}
The procedure by which target stars are selected in the simulation aims at maximizing the prospects for detecting the transits of small planets, and hence is mainly driven by stellar radius and apparent magnitude. In practice, this is done\footnote{The actual target selection procedure differs slightly from the one adopted in the simulation: stars will be selected for which a 2.25-$R_\earth$ planet can be detected in a single 4-hour transit at the $5\sigma$ level.} by determining whether a fiducial planet with an orbital period of 20 days could be detected by {\it TESS} transiting a given star. This results in a target star catalog that is approximately complete for short-period planets smaller than $2.25\,R_\earth$. From a stellar perspective, this also means that nearly all bright main-sequence stars with $T_{\rm eff}\!<\!6000\:{\rm K}$ are selected, while a decreasing fraction of hotter stars make it into the target star catalog. In effect, a limiting apparent magnitude $I_{\rm C}\!\la\!12$ is imposed for FGK dwarfs (cf.~fig.~17 of S15). Given this limiting apparent magnitude, virtually all main-sequence stars for which the detection of solar-like oscillations will be possible should already be included in the target star catalog (see Fig.~\ref{fig:stellar_tracks_120s_lum}).

Furthermore, according to fig.~16 of S15, a non-negligible\footnote{Using flicker measurements of 289 bright {\it Kepler} candidate exoplanet-host stars with $4500\:{\rm K}\!<\!T_{\rm eff}\!<\!6650\:{\rm K}$, \citet{Bastien14} found that a Malmquist bias is responsible for a contamination of the sample by evolved stars, being that nearly $50\,\%$ of those stars are in fact subgiants.} number of subgiants end up being selected as target stars, even though they are far from optimal for transiting planet detection. Once {\it Gaia} \citep{Gaia} parallaxes become available, we expect to have excellent knowledge of target stellar radii and that information could then be used to screen out, or else to deliberately target, subgiants. Here we advocate for the latter. As can be seen from Fig.~\ref{fig:stellar_tracks_120s_lum}, bright subgiants are attractive targets for the 2-min-cadence slots reserved for asteroseismology. In what follows, we assess the overall asteroseismic potential of subgiant stars and the resulting impact on the asteroseismic yield of target hosts. 

Having access to the all-sky bright catalog from where target stars have been selected, we made use of the known stellar properties to isolate all subgiant stars that fall into {\it TESS}'s field of view. These stars were then ranked in order of decreasing brightness and the detection test was run assuming they would be observed at the 2-min cadence. Simply ranking stars by brightness does not necessarily constitute the optimal procedure for selecting potential asteroseismic targets, as there is also a dependence of the detectability of solar-like oscillations on stellar mass and effective temperature along the subgiant branch (cf.~Fig.~\ref{fig:stellar_tracks_120s_lum}), not to mention the effect of the length of the observations. This simple approach is nonetheless suitable for arguing our point and allows as well setting an upper bound on the number of pixels required to accommodate these potential asteroseismic subgiants. 

\begin{figure}[!t]
\centering
  \subfigure{%
  \includegraphics[width=.5\textwidth,trim={1.25cm 0cm 1.5cm 2cm},clip]{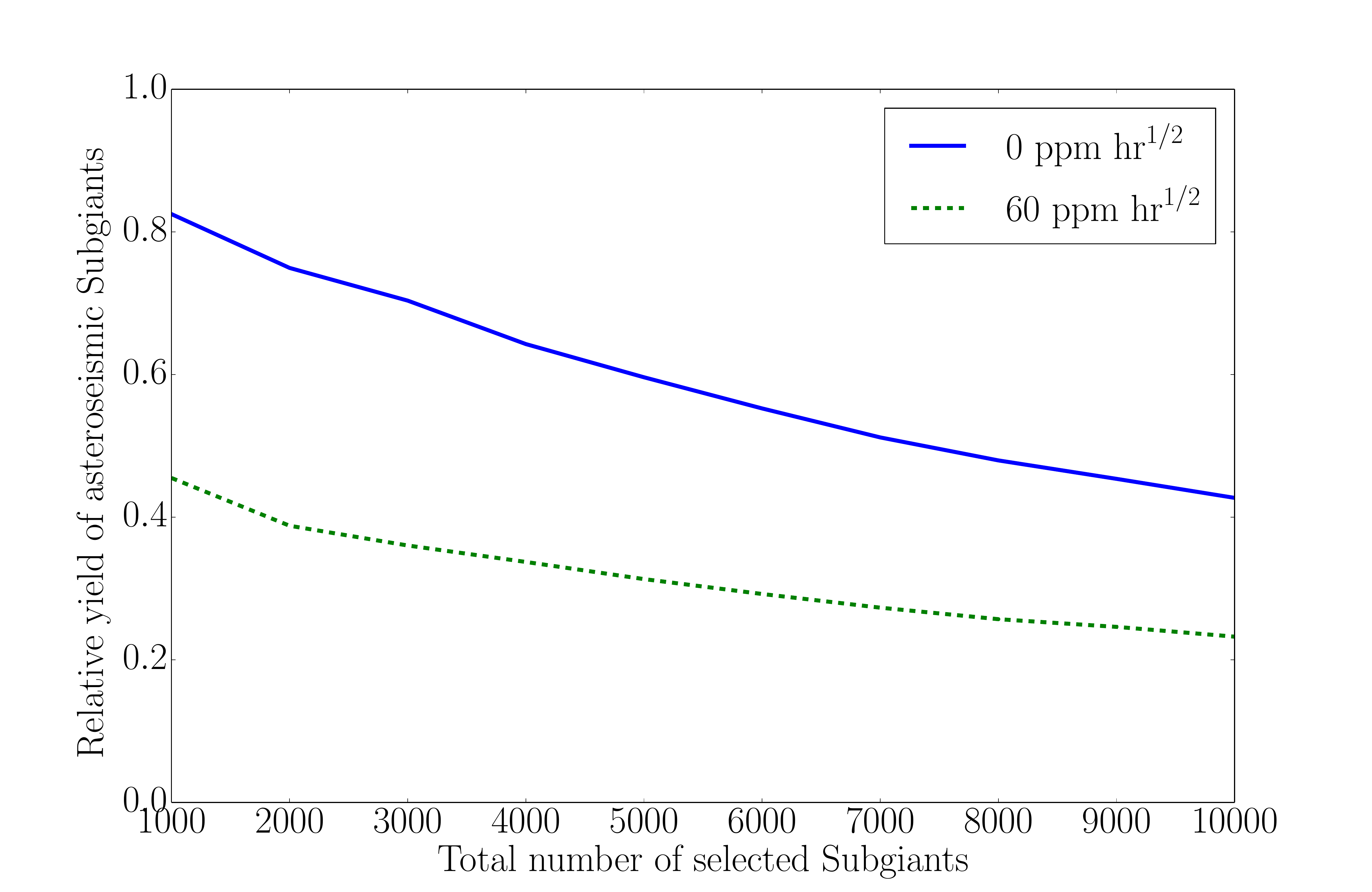}}\hfill
  \subfigure{%
  \includegraphics[width=.5\textwidth,trim={1.25cm 0cm 1.5cm 2cm},clip]{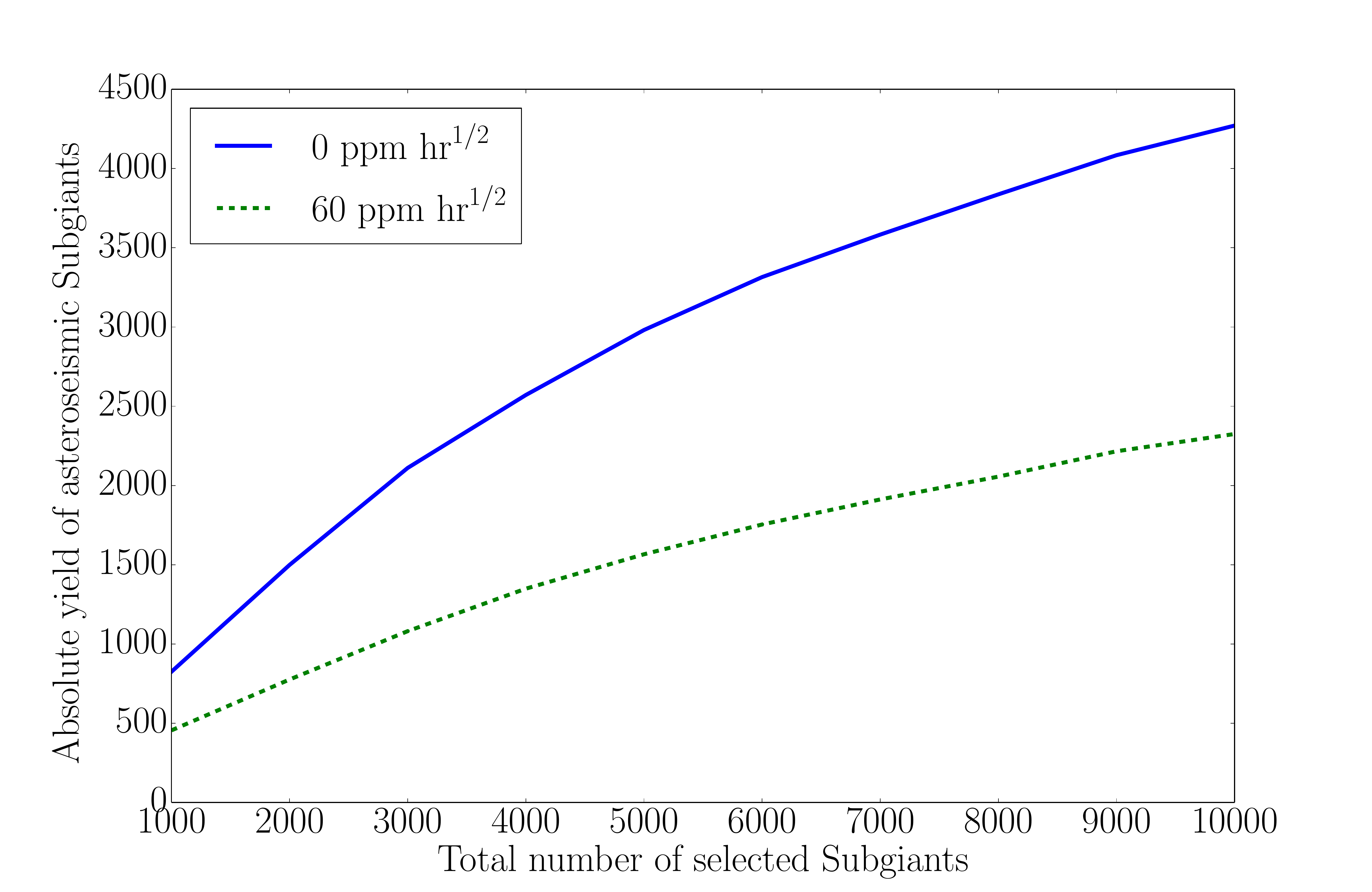}}\\
  \subfigure{%
  \includegraphics[width=.5\textwidth,trim={1.25cm 0cm 1.5cm 2cm},clip]{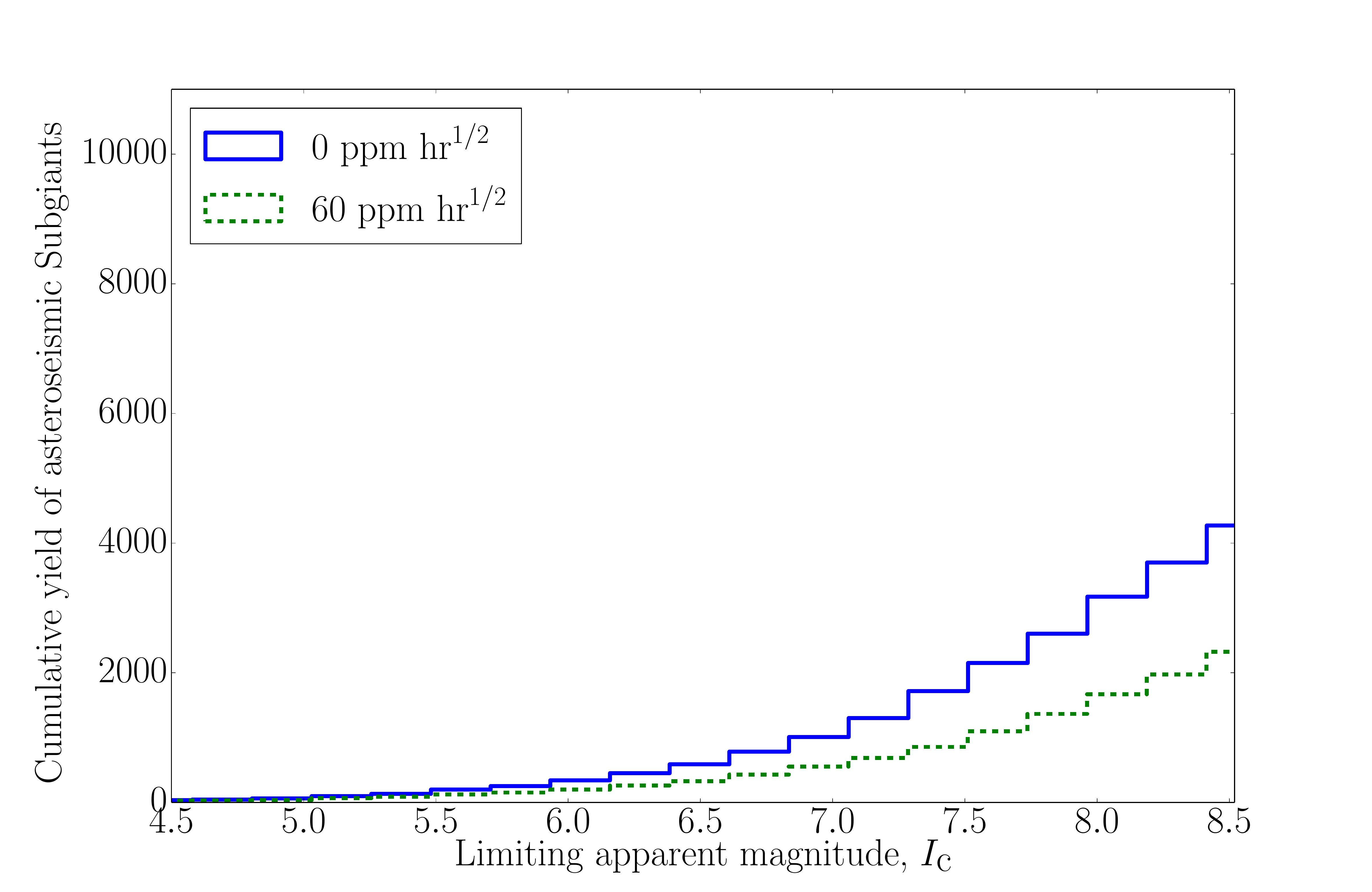}}\hfill
  \subfigure{%
  \includegraphics[width=.5\textwidth,trim={1.25cm 0cm 1.5cm 2cm},clip]{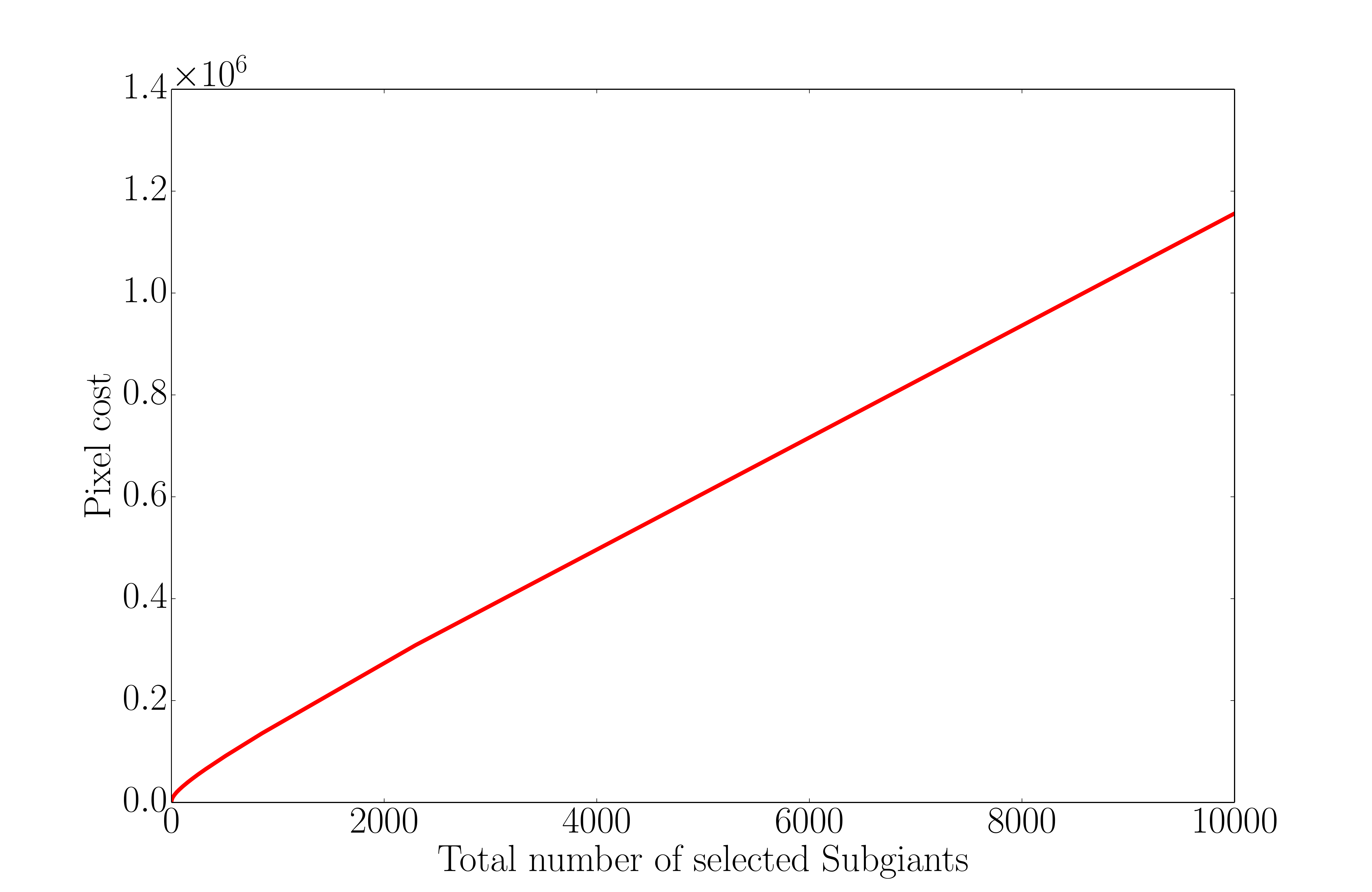}}
  \caption{\small Asteroseismic potential of subgiant stars. Top panels: Relative (left) and absolute (right) asteroseismic yield as a function of the total number of subgiants selected as target stars (ranked in order of decreasing brightness). Bottom left panel: Cumulative yield of asteroseismic subgiants as a function of limiting apparent magnitude. Bottom right panel: Cumulative pixel cost as a function of the total number of subgiants selected as target stars (ranked in order of decreasing brightness). Systems were assumed to be isolated (i.e., $D\!=\!1$). A systematic noise level of either $\sigma_{\rm sys}\!=\!0\:{\rm ppm\,hr^{1/2}}$ or $\sigma_{\rm sys}\!=\!60\:{\rm ppm\,hr^{1/2}}$ was considered, as indicated.}\label{fig:subgiants}
\end{figure}

Figure \ref{fig:subgiants} summarizes the outcome of this exercise. The horizontal axes in the top panels of Fig.~\ref{fig:subgiants} represent the total number of selected subgiants (after being ranked in order of decreasing brightness), with the vertical axes representing the relative (top left) and absolute (top right) yield of asteroseismic subgiants. The bottom left panel provides an alternative perspective, by plotting the cumulative yield of asteroseismic subgiants as a function of limiting apparent magnitude. The cumulative number of pixels in the target masks is shown in the bottom right panel. If, for instance, we were to select the brightest $1\!\times\!10^4$ ($5\!\times\!10^3$) subgiants in {\it TESS}'s field-of-view, one would be able to detect solar-like oscillations in $\sim\!43\,\%$ ($\sim\!60\,\%$) of those stars assuming a systematic noise level of $\sigma_{\rm sys}\!=\!0\:{\rm ppm\,hr^{1/2}}$. Furthermore, this would equate to a cumulative pixel cost of $\sim\!1.2\!\times\!10^6$ ($\sim\!6.1\!\times\!10^5$) pixels over the course of the mission or $\sim\!1.1\!\times\!10^4$ ($\sim\!5.8\!\times\!10^3$) pixels on average per camera for any given observation sector. For reference, due to onboard storage and bandwidth limitations, an allocation of $\sim\!1.4\:{\rm megapixels}$ per camera for all types of 2-min-cadence targets has been set as the design goal.

\begin{figure}[!t]
\centering
  \subfigure[$\sigma_{\rm sys}\!=\!0\:{\rm ppm\,hr^{1/2}}$.]{%
  \includegraphics[width=.48\textwidth,trim={4cm 0cm 7.25cm 0cm},clip]{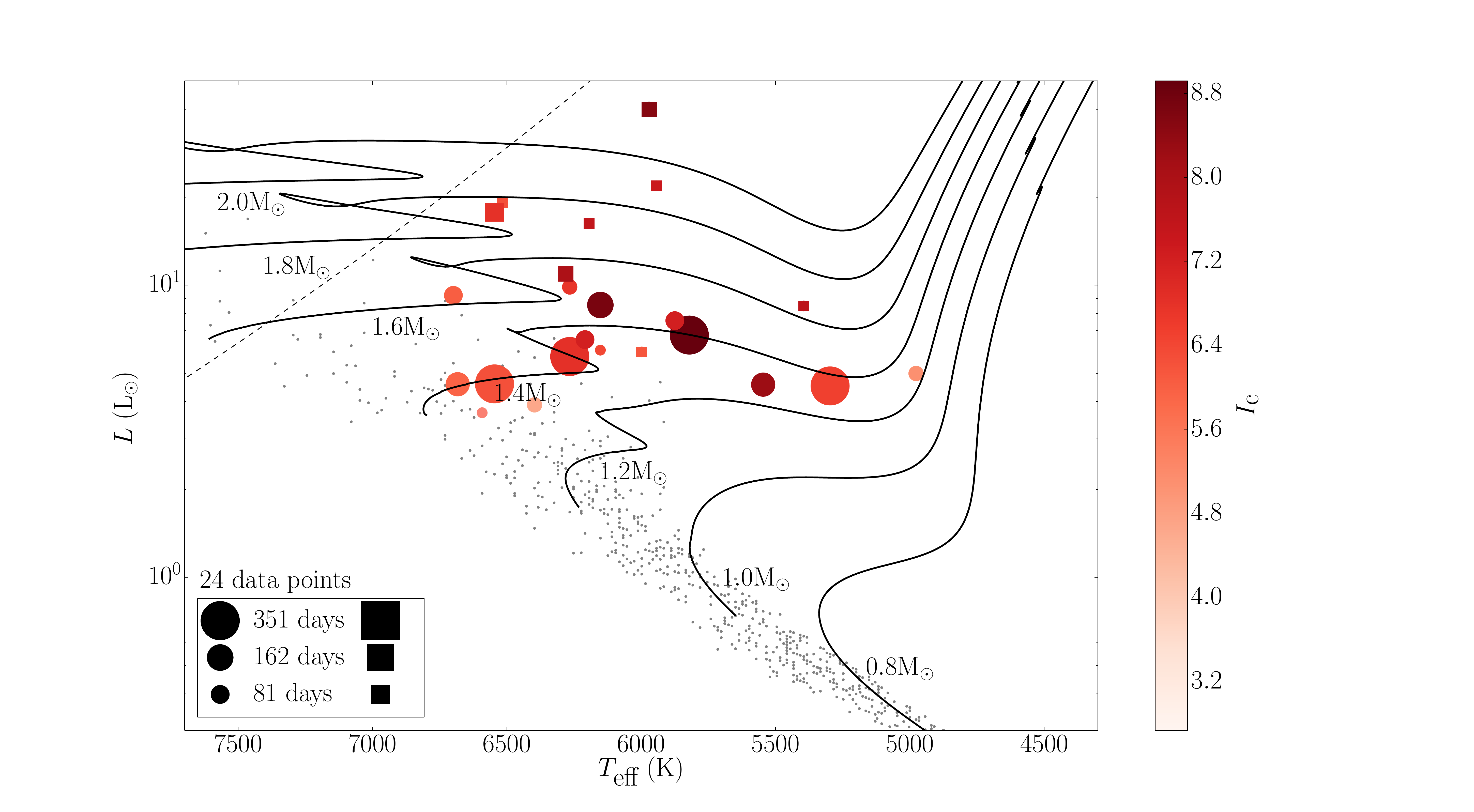}}\hfill
  \subfigure[$\sigma_{\rm sys}\!=\!60\:{\rm ppm\,hr^{1/2}}$.]{%
  \includegraphics[width=.48\textwidth,trim={4cm 0cm 7.25cm 0cm},clip]{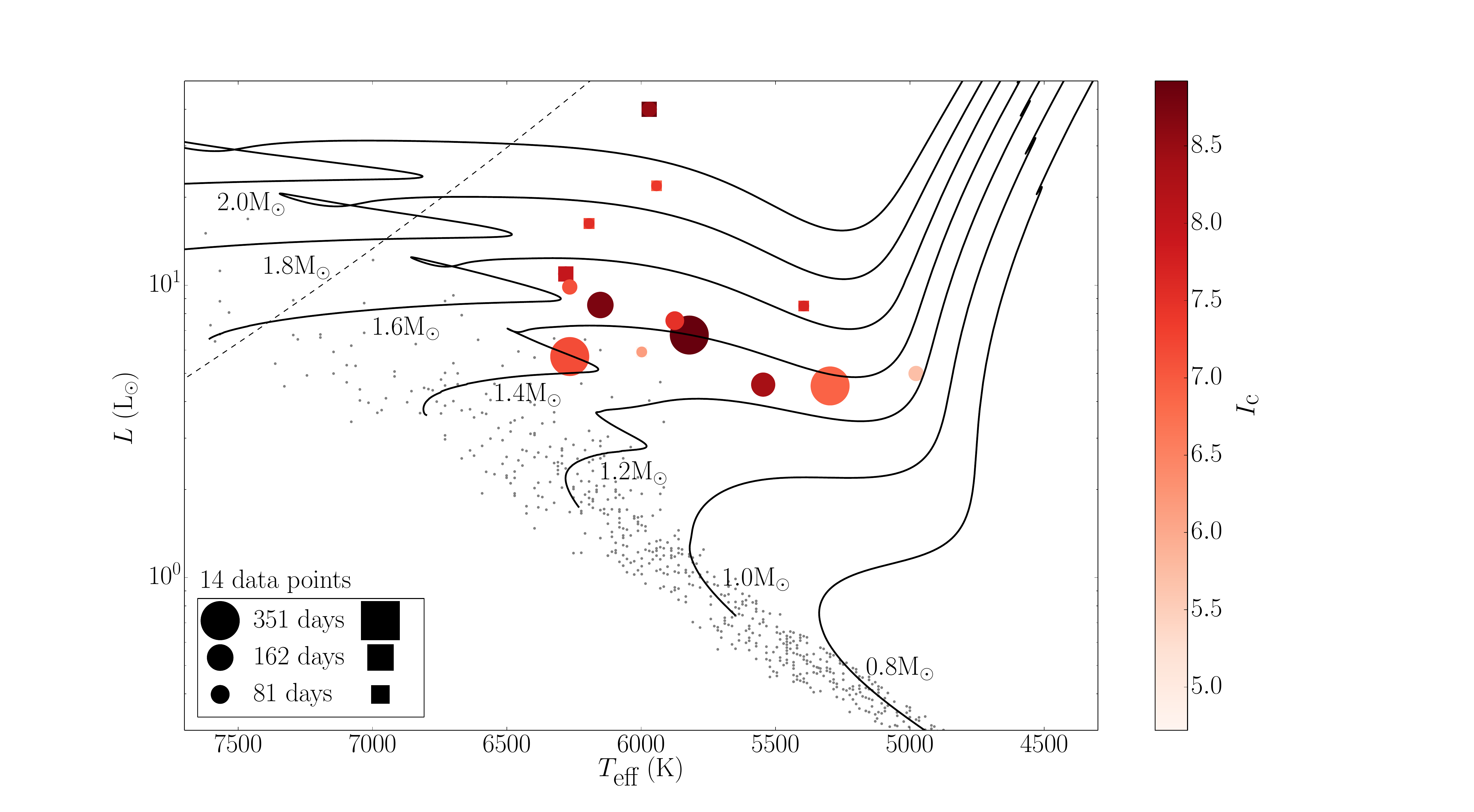}}
  \caption{\small Asteroseismic yield of exoplanet-host stars (target stars). The yield is computed for a single trial. Data points are color-coded according to apparent magnitude and their size is proportional to the observing length. Squares correspond to those extra stars with asteroseismic detections once the brightest subgiants have been included during target selection. Gray dots represent the underlying synthetic population of host stars from S15. Solar-calibrated evolutionary tracks spanning the mass range 0.8--$2.0\,{\rm M}_\sun$ (in steps of $0.2\,{\rm M}_\sun$) are shown as continuous lines. The slanted dashed line represents the red edge of the $\delta$ Scuti instability strip. A systematic noise level of either $\sigma_{\rm sys}\!=\!0\:{\rm ppm\,hr^{1/2}}$ or $\sigma_{\rm sys}\!=\!60\:{\rm ppm\,hr^{1/2}}$ was considered, as indicated.}\label{fig:targetstars}
\end{figure}

Let us then assume that we select the brightest $1\!\times\!10^4$ subgiants in {\it TESS}'s field-of-view and observe them at the 2-min cadence. What impact could this potentially have on the asteroseismic yield of target hosts? Doing this corresponds to setting a limiting apparent magnitude $I_{\rm C}\!\sim\!8.5$ (cf.~bottom left panel of Fig.~\ref{fig:subgiants}). We now apply this magnitude cut to the synthetic population of subgiant hosts in FFIs and run the detection test\footnote{The procedure described will in principle only provide a lower bound on the asteroseismic yield of subgiant hosts. The planet yield for FFI stars is estimated based on a 30-min cadence, which can smear out short-duration and/or high-impact-parameter transits. Were we to observe the brightest $1\!\times\!10^4$ subgiants in {\it TESS}'s field-of-view at a 2-min cadence, planets that would otherwise remain undetectable using the 30-min cadence could now in principle be detected. We tested this by seeding these bright subgiants with planets having adopted revised planet occurrence rates around evolved stars \citetext{Thomas S.~H.~North et al., submitted}, after which we simulated {\it TESS} observations at the 2- and 30-min cadences. The resulting lack of difference between the two planet yields (i.e., obtained for either cadence) can be understood in terms of the long transit durations of planets about large stars (with a mean duration of 18 hours for the detected planets in this exercise), so that switching from a 30- to a 2-min cadence does not lead to a significant improvement in the planet yield.}. Figure \ref{fig:targetstars} shows the asteroseismic yield of exoplanet-host target stars for a single representative trial. This is dominated by subgiant stars. We assume Poisson statistics in estimating the statistical uncertainties and obtain $24\pm5$ or $14\pm4$ host stars depending on whether $\sigma_{\rm sys}\!=\!0\:{\rm ppm\,hr^{1/2}}$ or $\sigma_{\rm sys}\!=\!60\:{\rm ppm\,hr^{1/2}}$ (to be compared to $16\pm4$ or $8\pm3$ before inclusion of the brightest subgiants). For intermediate values of $\sigma_{\rm sys}$, the yield can be simply estimated by linear interpolation.

We note that this yield may be affected by biases in the planet occurrence rates upon which the simulation is based. S15 point out that such biases may be as high as $\sim\!40\,\%$ across all planetary sizes and periods. We further note that the adopted occurrence rates do not account for the expected effects of post-main-sequence evolution on the occurrence of planets migrating into close-in orbits \citep[e.g.,][]{Frewen16}.

\begin{figure}[!t]
\centering
  \subfigure[$\sigma_{\rm sys}\!=\!0\:{\rm ppm\,hr^{1/2}}$.]{%
  \includegraphics[width=.48\textwidth,trim={4cm 0cm 7.25cm 0cm},clip]{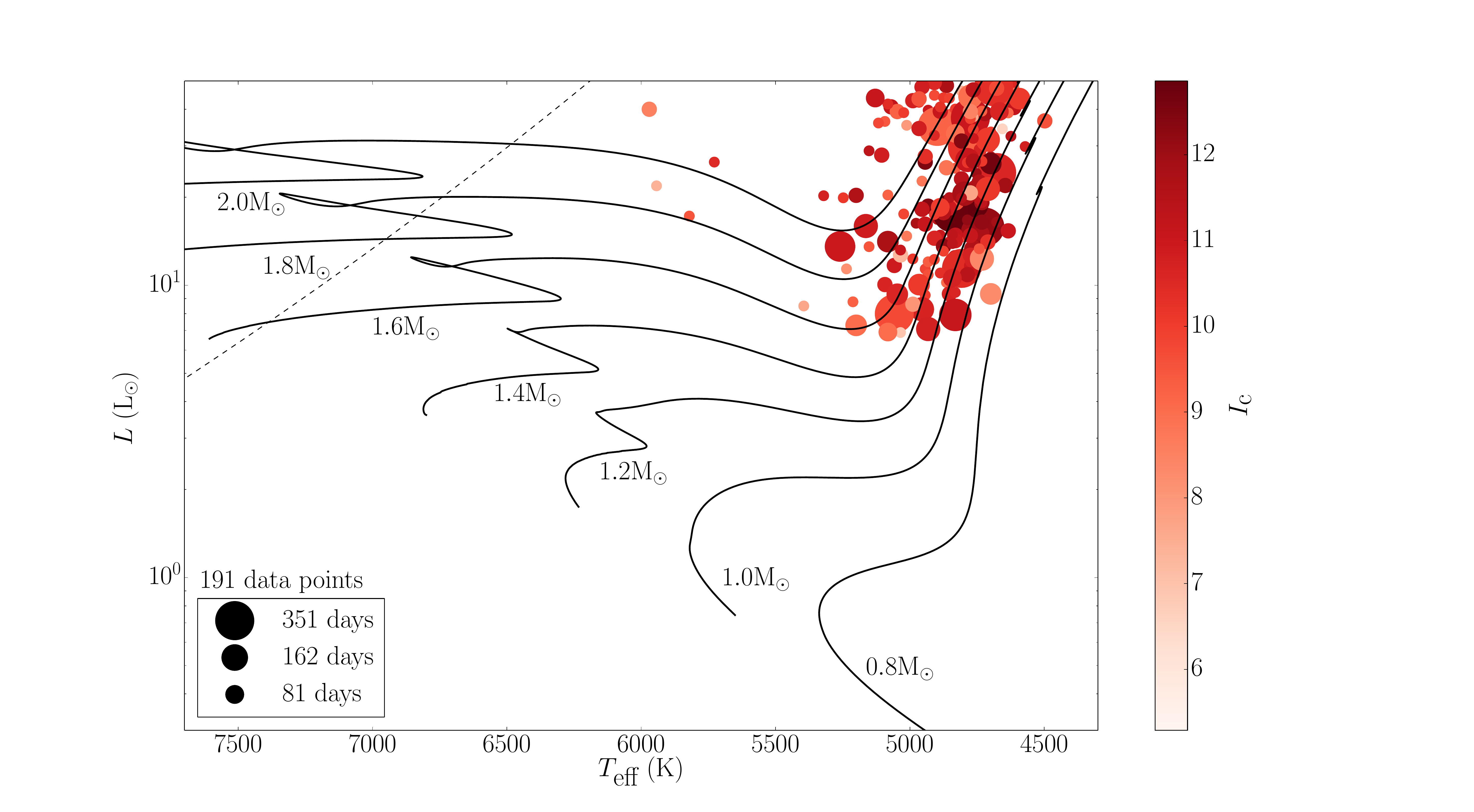}}\hfill
  \subfigure[$\sigma_{\rm sys}\!=\!60\:{\rm ppm\,hr^{1/2}}$.]{%
  \includegraphics[width=.48\textwidth,trim={4cm 0cm 7.25cm 0cm},clip]{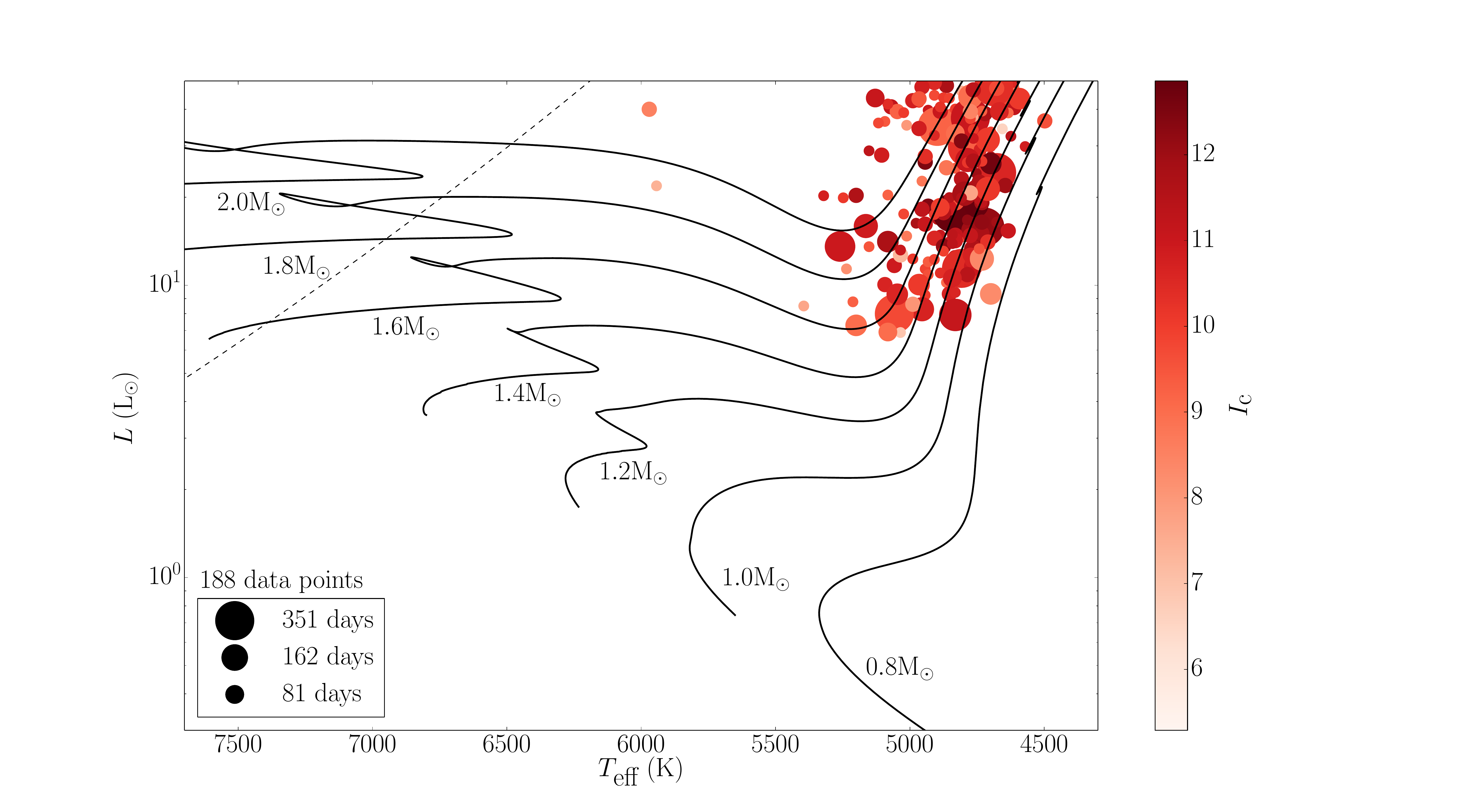}}
  \caption{\small Asteroseismic yield of exoplanet-host stars (FFI stars). The yield is computed for a single trial. Data points are color-coded according to apparent magnitude and their size is proportional to the observing length. Solar-calibrated evolutionary tracks spanning the mass range 0.8--$2.0\,{\rm M}_\sun$ (in steps of $0.2\,{\rm M}_\sun$) are shown as continuous lines. The slanted dashed line represents the red edge of the $\delta$ Scuti instability strip. A systematic noise level of either $\sigma_{\rm sys}\!=\!0\:{\rm ppm\,hr^{1/2}}$ or $\sigma_{\rm sys}\!=\!60\:{\rm ppm\,hr^{1/2}}$ was considered, as indicated.}\label{fig:ffistars}
\end{figure}

\subsection{{\it TESS} FFI hosts}\label{sec:simyield30min}
Figure \ref{fig:ffistars} shows the asteroseismic yield of exoplanet-host FFI stars for a single representative trial. The depicted host stars are in their vast majority low-luminosity red giants. Assuming Poisson statistics, we obtain $191\pm14$ or $188\pm14$ host stars depending on whether $\sigma_{\rm sys}\!=\!0\:{\rm ppm\,hr^{1/2}}$ or $\sigma_{\rm sys}\!=\!60\:{\rm ppm\,hr^{1/2}}$. We note that the adopted occurrence rates \citep[from][for $T_{\rm eff}\!>\!4000\:{\rm K}$]{Fressin13} do not account for physical and orbital changes of planets as their parent stars evolve off the main sequence. Such evolutionary effects might be substantial, as there seem to be fewer close-in giant planets around evolved stars than main-sequence stars \citep[e.g.,][]{Bowler10,JJ10}. An investigation of evolutionary effects on planet occurrence rates, and hence on {\it TESS} planet yields, is beyond the scope of this work. Since in S15 at least 2 transits need to be observed for a planet to be flagged as detectable, the yield shown in Fig.~\ref{fig:ffistars} does not take into account single-transit events associated with long-period planets, which can be followed up with radial-velocity (RV) observations in order to characterize the planet \citep[e.g.,][]{YeeGaudi}. Given the large expected yield of red-giant stars with detectable solar-like oscillations, it is likely that there will be a significant number of such single-transit events around asteroseismic hosts.

\begin{figure}[!t]
\centering
\includegraphics[width=0.8\linewidth]{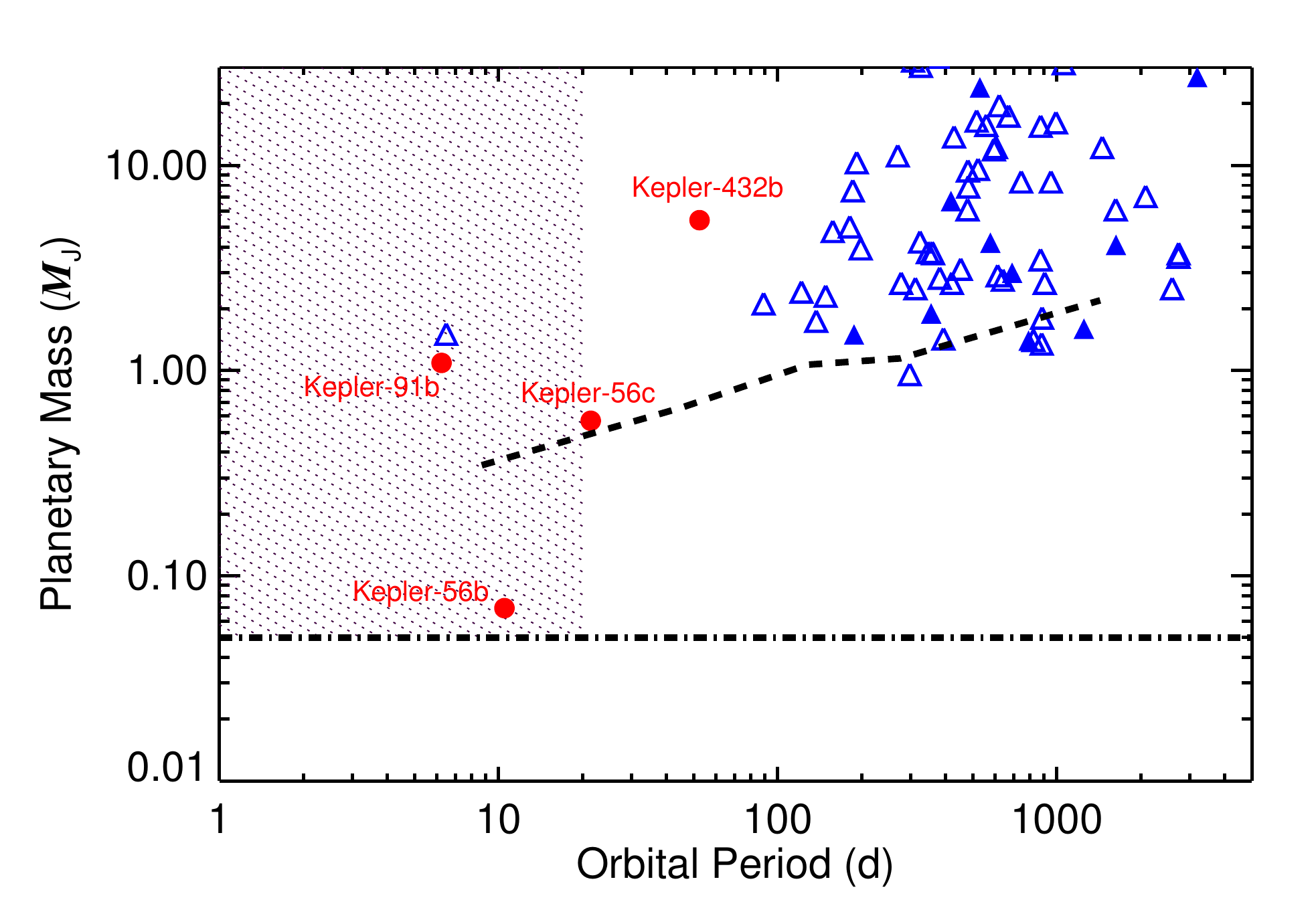}
\caption{\small Mass-period diagram of known exoplanets orbiting red-giant-branch stars. Planets detected by the transit method are depicted as red circles and those detected in RV surveys as blue triangles (open triangles correspond to mean planetary masses assuming random orbital orientations). The dashed line represents the median RV detection threshold for mean masses from \citet{Bowler10}. The dashed-dotted line marks the mass of Neptune and represents an approximate {\it TESS} detection limit. The shaded area approximately corresponds to the parameter space that will be probed by {\it TESS}.\label{fig:rgplanets}}
\end{figure}

Shown in Fig.~\ref{fig:rgplanets} is a mass-period diagram of known exoplanets orbiting red-giant-branch stars \citep[adapted from][]{Huber15}. Despite the dearth of close-in giants planets (with $M_{\rm p}\!\ga\!0.5\,M_{\rm J}$) unveiled by RV surveys \citep[e.g.,][]{Johnson07}, data from {\it Kepler} have led to the discovery of several giant planets with short orbital periods ($P_{\rm orb}\!\la\!50\:{\rm d}$) orbiting asteroseismic red-giant branch stars \citep[4 planets in 3 systems, to be precise;][]{Kepler-56,Kepler-91,Kepler-432,Quinn15}. The latter may be hinting at the existence of a population of warm sub-Jovian planets around evolved stars that has remained elusive to RV surveys. The shaded area in Fig.~\ref{fig:rgplanets} approximately corresponds to the parameter space that will be probed by {\it TESS}, which will be mainly sensitive to planets with orbital periods\footnote{A fiducial planet with an orbital period of 13 days in a circular orbit around a low-luminosity red giant will have $a/R\!\sim\!5$, where $a$ is the semimajor axis, hence well above the Roche limit.} $P_{\rm orb}\!\la\!20\:{\rm d}$. Such parameter space is inaccessible to RV surveys at the low planetary-mass range. 

\section{Asteroseismic yield of confirmed exoplanet-host stars}\label{sec:realyield}
We are now interested in assessing {\it TESS}'s asteroseismic yield of known (i.e., confirmed) exoplanet-host stars, assuming these will all be selected as target stars. We used the NASA Exoplanet Archive\footnote{\url{http://exoplanetarchive.ipac.caltech.edu/}} \citep[][]{Akeson13} to identify all known host stars (1182 at the time of writing after discarding the few known circumbinary planetary systems). The minimum amount of information on a given star that must be available in order to compute the probability of detecting solar-like oscillations comprises its celestial coordinates, $I_{\rm C}$-band magnitude, $T_{\rm eff}$ and $R$ (we henceforth enforce the simplifying assumption that stars are isolated, i.e., $D\!=\!1$). While celestial coordinates are readily available for all known hosts, the same is not true for the remaining three quantities, and we will often need to derive them based on ancillary stellar properties. We started by grouping the known host stars according to the set of available properties, as follows: 
\begin{enumerate}
\item Stars with an entry in the {\it Hipparcos} catalog. For the known hosts with an entry in the Extended {\it Hipparcos} Compilation \citep[XHIP;][]{XHIP}, $I_{\rm C}$-band magnitudes are readily available. Whenever available in the Exoplanet Archive, $T_{\rm eff}$ and/or $R$ values were used. When not available, these then had to be derived. The effective temperature was calculated using the $(B-V)$-$T_{\rm eff}$ relation from \citet{Torres10}, which uses the $B-V$ color index as input. In order to compute the stellar radius, the stellar luminosity was first calculated via the {\it Hipparcos} parallax, $\pi$, using \citep{Pijpers03}:
\begin{equation}
\label{eq:Lum}
\log(L/{\rm L}_\sun) = 4.0 + 0.4M_{\rm bol,\sun} - 2.0\log\pi[{\rm mas}] - 0.4(V - A_V + {\rm BC}_{V}) \, ,
\end{equation}  
where we have adopted $M_{\rm bol,\sun}\!=\!4.73\:{\rm mag}$ \citep{Torres10} for the bolometric magnitude of the Sun, $V$ is the apparent visual magnitude (available in XHIP), $A_V$ is the reddening (assumed negligible for the bright stars in question), and ${\rm BC}_{V}$ are the bolometric corrections from the \citet{Flower96} polynomials presented in \citet{Torres10}, which use $T_{\rm eff}$ as input. Stellar radii were then computed by rearranging Stefan--Boltzmann's law. Only stars with fractional parallax errors smaller than $25\,\%$ were retained. A total of 385 stars fell under this group.
\item $I_{\rm C}$-band magnitude, $T_{\rm eff}$ and $R$ directly available from the Exoplanet Archive. These were used in the case of 33 host stars.
\item No available $I_{\rm C}$-band magnitude. For the numerous {\it Kepler} and {\it K2} hosts, estimates of $T_{\rm eff}$ and $R$ are generally available in the Exoplanet Archive, but an estimate of $I_{\rm C}$ is usually not. In such cases, we start by computing the Johnson--Cousins $R-I_{\rm C}$ color index from 2MASS $JHK_S$ colors on the main sequence \citep{Bilir08}:
\begin{equation}
\label{eq:RIcolor}
R-I_{\rm C} = 0.954(J-H) + 0.593(H-K_S) + 0.025 \, .
\end{equation}  
The previous equation is then used in combination with the Johnson--Cousins $UBVRI_{\rm C}$ to SDSS $ugriz$ transformations from \citet{Jordi06}, to give $I_{\rm C}$ in terms of 2MASS $JHK_S$ and SDSS $r$ photometry, i.e.,
\begin{equation}
\label{eq:Iband}
I_{\rm C} = r - 1.239(R-I_{\rm C}) - 0.104 \, .
\end{equation}
This enabled us to gather all input quantities needed to run the detection test for 362 {\it Kepler} and {\it K2} hosts. Alternatively, for other families of hosts the $I_{\rm C}$-band magnitude could be estimated based on the statistical color-color relation of \citet{Caldwell93} provided $B-V$ and $V$ are available (with separate sets of coefficients tabulated according to luminosity class). Further requiring that $R$ is available (since $T_{\rm eff}$ could always be estimated from the $B-V$ color index), this ended up providing all input quantities for an additional 182 hosts. We note that for 133 of these stars we had to rely on the properties available through the Exoplanet Orbit Database\footnote{\url{http://www.exoplanets.org/}} \citep{EOD}.     
\end{enumerate}  

\begin{figure}[!t]
\centering
  \subfigure[$\sigma_{\rm sys}\!=\!0\:{\rm ppm\,hr^{1/2}}$.]{%
  \includegraphics[width=.48\textwidth,trim={4cm 0cm 7.25cm 0cm},clip]{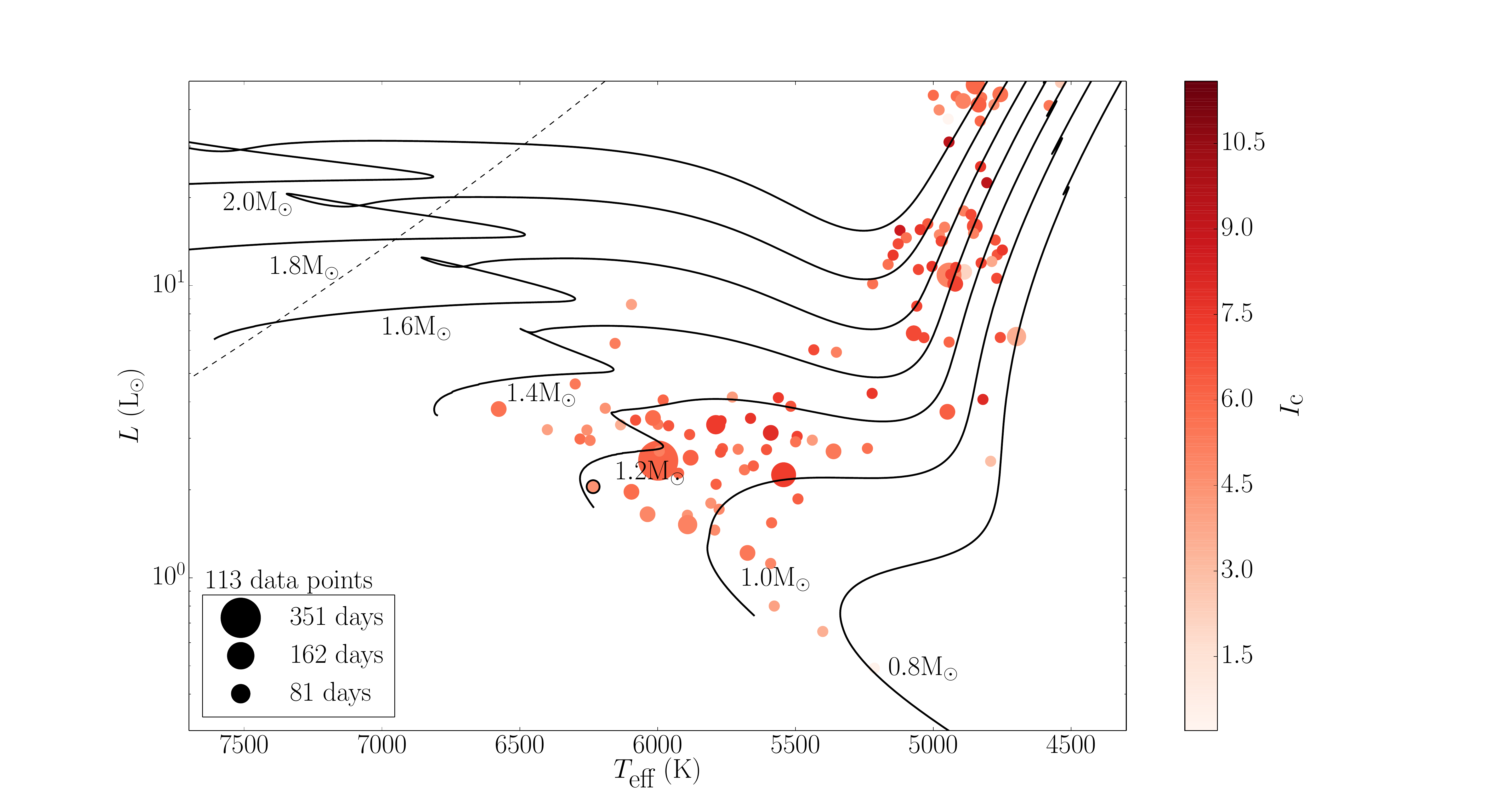}}\hfill
  \subfigure[$\sigma_{\rm sys}\!=\!60\:{\rm ppm\,hr^{1/2}}$.]{%
  \includegraphics[width=.48\textwidth,trim={4cm 0cm 7.25cm 0cm},clip]{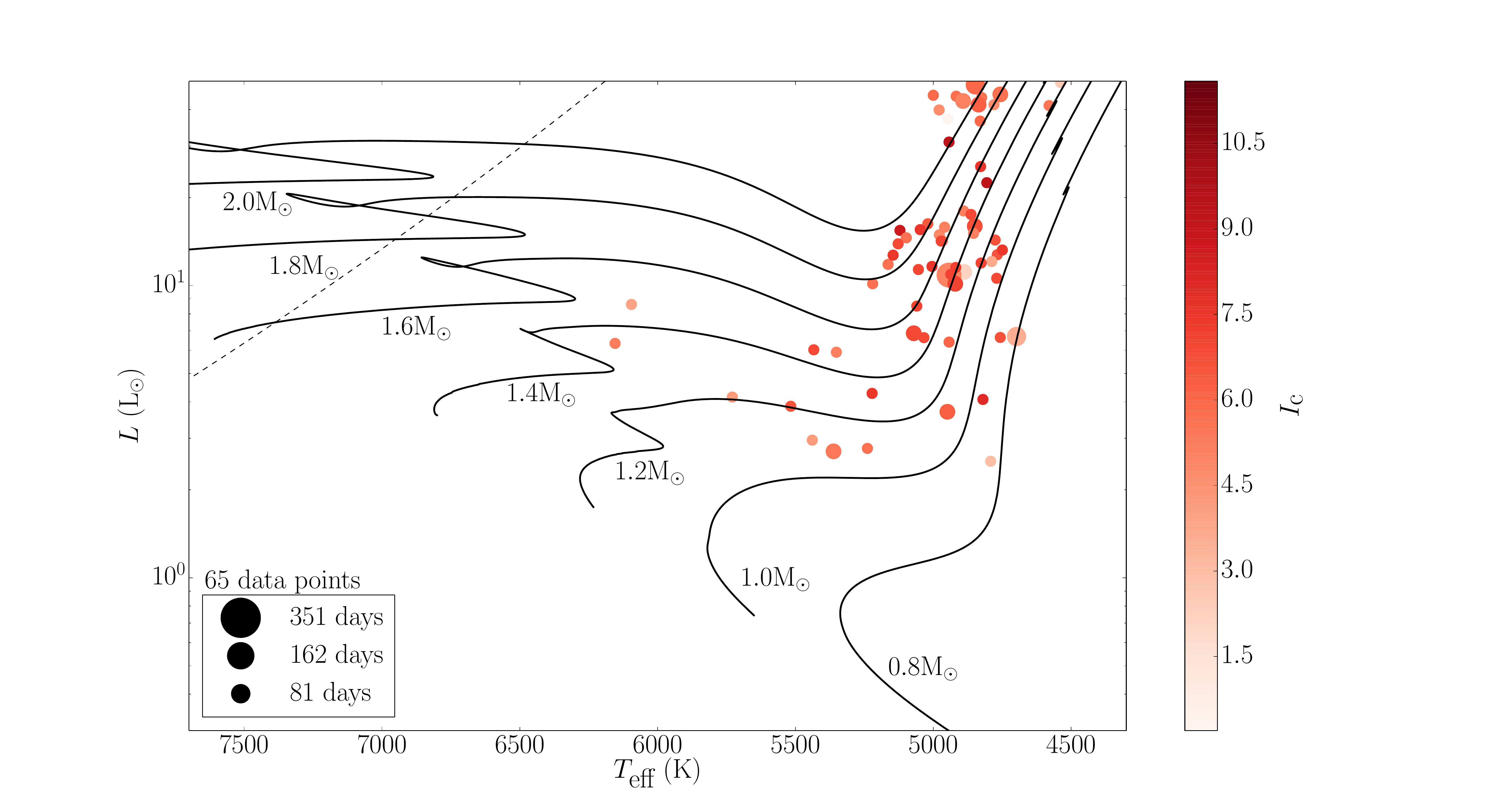}}
  \caption{\small Asteroseismic yield of known exoplanet-host stars for a cadence of $\Delta t\!=\!2\:{\rm min}$. Data points are color-coded according to apparent magnitude and their size is proportional to the observing length. All but one system had their first planet (i.e., with suffix ``b'') detected using RV measurements. The only non-RV host had its first planet detected through direct imaging (represented by a circle with a black rim in the left panel). Solar-calibrated evolutionary tracks spanning the mass range 0.8--$2.0\,{\rm M}_\sun$ (in steps of $0.2\,{\rm M}_\sun$) are shown as continuous lines. The slanted dashed line represents the red edge of the $\delta$ Scuti instability strip. A systematic noise level of either $\sigma_{\rm sys}\!=\!0\:{\rm ppm\,hr^{1/2}}$ or $\sigma_{\rm sys}\!=\!60\:{\rm ppm\,hr^{1/2}}$ was considered, as indicated.}\label{fig:yieldknown}
\end{figure}

Stars which do not fall into one of the groups above were discarded. There are 962 known hosts for which all the relevant input quantities are available. Of these, 832 occupy that portion of the H--R diagram populated by solar-type and (low-luminosity) red-giant stars, and for which we ran the detection test. Figure \ref{fig:yieldknown} shows the asteroseismic yield of known exoplanet-host stars assuming either $\sigma_{\rm sys}\!=\!0\:{\rm ppm\,hr^{1/2}}$ or $\sigma_{\rm sys}\!=\!60\:{\rm ppm\,hr^{1/2}}$. For intermediate values of $\sigma_{\rm sys}$, the yield can again be estimated by linear interpolation. By considering the faster-than-standard 20-sec cadence, we may still expect to detect solar-like oscillations in a few extra high-$\nu_{\rm max}$ hosts. Allocation of these slots will only be relevant for stars with $\nu_{\rm max}$ larger than $\nu_{\rm Nyq,target}/2\!\sim\!2084\:{\rm \mu Hz}$, for which the attenuation factor, $\eta^2(\nu_{\rm max})$, exceeds $\sim\!20\,\%$. 

We remind the reader that the actual pointing coordinates will depend on the spacecraft's launch date. The yield, however, remains virtually unchanged if we were to adopt different pointing coordinates. Furthermore, we notice how the asteroseismic yield of known exoplanet-host stars is an order of magnitude greater than that of target hosts (cf.~left panels of Figs.~\ref{fig:targetstars} and \ref{fig:yieldknown}). This is simply the result of a selection effect. Firstly, {\it TESS} target stars are preferentially bright main-sequence stars with spectral types F5 and later, thus maximizing the prospects for detecting the transits of small planets. Secondly, {\it TESS} target hosts are restricted to transiting systems with short orbital periods, whereas known hosts are in their vast majority RV systems (hence allowing for a range of orbital inclinations) whose planets span a wider range in terms of orbital period (the median orbital period of planet ``b'' around main-sequence hosts in the left panel of Fig.~\ref{fig:yieldknown} is 480.3 days).

With over 100 solar-type and red-giant known hosts with detectable solar-like oscillations, this represents an invaluable stellar sample. The impact of having additional constraints from {\it TESS} asteroseismology on the characterization of known exoplanet-host stars, and consequently of their planetary systems, remains to be fully assessed. Also, we note that all but one system in Fig.~\ref{fig:yieldknown} were discovered using RV measurements and hence will be potential prime targets for the upcoming ESA {\it CHaracterising ExOPlanet Satellite} \citep[{\it CHEOPS};][]{CHEOPS}. {\it CHEOPS} will be monitoring bright ($V\!<\!12$) known hosts anywhere in the sky for transiting planets. Consequently, {\it TESS} could be providing asteroseismic measurements for a significant number of potential {\it CHEOPS} targets, a link that is yet to be explored.

\section{Summary and discussion}\label{sec:summary}
We have developed a simple test to estimate the detectability of solar-like oscillations in {\it TESS} photometry of any given star (Sect.~\ref{sec:detect_test}). The detection test looks for signatures of the bell-shaped power excess due to the oscillations. We applied the detection test along stellar-model tracks spanning a range of masses in order to predict the detectability of solar-like oscillations across the H--R diagram (Sect.~\ref{sec:detect_HR}). 

Detection of the power excess due to the oscillations as considered here, and hence the ability to measure $\nu_{\rm max}$, will generally mean that the large frequency separation $\Delta\nu$ can be readily extracted. Fundamental stellar properties can be estimated by comparing these two global asteroseismic parameters and complementary spectroscopic observables to the outputs of stellar evolutionary models. This so-called grid-based approach to the determination of stellar properties is currently well established \citep[e.g.,][]{StelloGrid,Basu10,Basu12,Creevey12}. A systematic study of {\it Kepler} planet-candidate hosts using asteroseismology was performed by \citet{HuberKOIs}, in which fundamental properties were determined for 66 host stars (with typical uncertainties of $3\,\%$ and $7\,\%$ in radius and mass, respectively) based on their average asteroseismic parameters. A similar approach was followed by \citet{Chaplin14} in estimating the fundamental properties of more than 500 main-sequence and subgiant field stars that had been observed for one month each with {\it Kepler}. For a subset of 87 of those stars, for which spectroscopic estimates of $T_{\rm eff}$ and metallicity were available, the median uncertainties obtained were $2.2\,\%$ in radius and $5.4\,\%$ in mass, with $57\,\%$ of the stars having age uncertainties smaller than $1\:{\rm Gyr}$. An outlook on the precision achievable by {\it TESS} on the estimation of stellar properties for a fiducial low-luminosity red giant is given in \citet{DaviesMiglio}. 

Furthermore, novel strategies have been developed that allow determining the stellar surface gravity for large samples of stars by directly measuring the amplitude of the brightness variations due to granulation and acoustic oscillations in the light curves \citep{flicker,Kallinger16}. However, owing to the shorter duration of {\it TESS} time series compared to {\it Kepler}'s and the fact that the instrumental/shot noise is now expected to dominate over granulation (cf.~Fig.~\ref{fig:Btot}), the robustness of such techniques when applied to {\it TESS} photometry remains to be tested. We have not addressed this issue here.

Based on an existing all-sky stellar and planetary synthetic population, we predicted the asteroseismic yield of the {\it TESS} mission, placing emphasis on the yield of exoplanet-host stars for which we expect to detect solar-like oscillations. This was done for both the target hosts (Sect.~\ref{sec:simyield2min}) and the full-frame-image or FFI hosts (Sect.~\ref{sec:simyield30min}). We predict that asteroseismology will become possible for a few dozen target hosts (mainly subgiant stars but also for a smaller number of F dwarfs) and for up to 200 FFI hosts (at the low-luminosity end of the red-giant branch). We also conducted a similar exercise based on a compilation of known host stars (Sect.~\ref{sec:realyield}), with the prediction being that over 100 solar-type and red-giant known hosts will have detectable solar-like oscillations. Altogether, this equates to a threefold improvement in the asteroseismic yield of exoplanet-host stars when compared to {\it Kepler}'s.

In Sect.~\ref{sec:simyield2min} we further advocate for the inclusion of as many bright subgiants as possible in the 2-min-cadence slots reserved for asteroseismology, where we assess the overall asteroseismic potential of subgiant stars and the resulting impact on the asteroseismic yield of target hosts. We should be able to use parallaxes from the ongoing {\it Gaia} mission to deliberately target these bright subgiants. More generally, {\it Gaia}-derived  luminosities could be used as strong constraints on the asteroseismic modeling, which should help improve the accuracy of the inferred stellar properties, in particular the stellar age.

\acknowledgments
The authors acknowledge the support of the UK Science and Technology Facilities Council (STFC). Funding for the Stellar Astrophysics Centre is provided by The Danish National Research Foundation (Grant DNRF106). D.H.~acknowledges support by the Australian Research Council's Discovery Projects funding scheme (project number DE140101364) and support by the National Aeronautics and Space Administration under Grant NNX14AB92G issued through the {\it Kepler} Participating Scientist Program. This research has made use of the NASA Exoplanet Archive, which is operated by the California Institute of Technology, under contract with the National Aeronautics and Space Administration under the Exoplanet Exploration Program. This research has made use of the Exoplanet Orbit Database and the Exoplanet Data Explorer at \url{exoplanets.org}.

{\it Facility:} \facility{{\it TESS}}

\bibliographystyle{apj}
\bibliography{biblio}

\end{document}